\documentclass[onecolumn,prb]{revtex4} 

\usepackage{epsfig}
\usepackage{amssymb}

\begin{document}

\def\k {{\bf k}}
\def\n {{\hat{\bf n}}}
\def\r {{\bf r}}
\def\u {{\bf u}}

\def\D {{\bf D}}
\def\G {{\bf G}}
\def\H {{\bf H}}
\def\A {{\bf A}}

\title{ Vector vibrations and the Ioffe-Regel crossover 
in disordered lattices}

\author{S.~N.~Taraskin and S.~R.~Elliott} 
\affiliation{Department of Chemistry, University of Cambridge,
             Lensfield Road, Cambridge CB2 1EW, UK}

\date{\today}

\begin{abstract}
The spectral density for vector vibrations in 
the f.c.c. lattice with force-constant disorder 
is analysed within the coherent potential approximation. 
The phase diagram showing the weak- and strong-scattering 
regimes is presented and compared with that  for electrons. 
The weak-scattering regime for external long-wavelength 
vibrational plane waves is shown to be due to 
sum-rule correlations in the dynamical matrix. 
A secondary peak below the Brillouin peak for sufficiently 
large wavevectors is found for the lattice models. 
The results obtained are supported by precise 
numerical solutions. 
\end{abstract}

\maketitle


%
%
%
%
\section{Introduction}
\label{s0}

Vibrational excitations in disordered structures 
is an active field of investigation by different methods. 
The first approaches were developed  
many decades ago 
for the case of substitutional alloys (see e.g. 
Ref.~\cite{Lifshitz_64,Taylor_67,Flicker_73,Elliott_74} 
and references therein). 
Disorder was introduced there 
by the random positioning of two or more  
atomic species (of different masses) onto 
lattice sites. 
Successful analytical methods have also been adapted  
for such problems 
\cite{Elliott_74,Economou_83:book,Gonis_92:book}. 
The rigidity-percolation problem 
has been solved to some extent numerically and 
within the mean-field approximations 
\cite{Feng_85,Garboczi_85:VDOS}. 
Atomic vibrations in structures with  fractal geometries were 
investigated in Refs.~\cite{Alexander_82,Bunde_92}. 
Some  phenomenological
\cite{Galperin_89:review,Klinger_88:review,Buchenau_92,Heuer_98} 
and more realistic 
\cite{Allen_93,Feldman_93,Fabian_99}
 models of atomic 
vibrations in glasses have been developed 
in order to describe the thermodynamical and 
phonon-transport 
peculiarities of such materials. 
Various computer experiments 
have also been performed on disordered structures 
(see e.g. \cite{Taraskin_99:PRB} and references therein). 

Recently, the methods of the mode-coupling theory 
have been applied to investigate short-time 
vibrational dynamics, especially in the low-energy 
regime \cite{Gotze_00}. 
The low-energy behaviour of the vibrational density 
of states and dynamical structure factor 
have been investigated in disordered lattice models 
within the coherent potential 
approximation (CPA) 
\cite{Schirmacher_98,Martin_Mayor_00,Taraskin_01:PRL}, 
and in positionally disordered models 
treated  perturbatively 
in the low inverse-particle-density limit 
\cite{Mezard_99,Martin_Mayor_01:JCP,Grigera_01,Grigera_01:BP}.  
The spectral properties of the Euclidian random matrices 
associated with topologically 
disordered systems \cite{Mezard_99} have been 
investigated within a supersymmetric 
statistic field theory in Ref.~\cite{Offer_00}. 
The spectral properties of the Laplacian defined on 
 a random graph \cite{Biroli_99} and 
''small-world'' lattices 
\cite{Monasson_99} are related to topologically disordered 
solids and have  also been  investigated. 

Previously, vibrational displacements in 
disordered structures have been  
treated as scalars,  which is a convenient simplification 
of the vector nature of atomic vibrations 
(although vector vibrations were considered in the 
rigidity-percolation problem \cite{Feng_85}).  
However, a consistent treatment of vector 
 vibrations in disordered lattices has 
not been performed up to now. 
The main purpose of this paper is to fill this gap 
and to give a mean-field description of   
vector vibrational models, namely: 
(i) to derive the CPA equations for vector vibrations in 
disordered lattices with force-constant disorder; 
(ii) to solve them analytically in limiting cases and numerically 
through the whole energy range; 
(iii) to apply the solution to investigate  the scattering 
properties of  external plane-wave 
vibrational excitations and, in particular, 
the Ioffe-Regel crossover from a weak- to 
strong-scattering regime; 
(iv) to demonstrate the importance of  
sum-rule correlations in the dynamical matrix 
for scattering properties of vibrational external plane waves.  
One of the  main results of the paper is the 
phase diagram for vector vibrations in the 
representative f.c.c. lattice showing 
the ranges of weak- and strong-scattering regimes 
for different degrees of disorder. 
Another result concerns a comparative analysis 
of the vibrational problem with the electronic one,  
and an explanation of the essential differences in phase 
diagrams 
for these problems. 
All the major  derivations and conclusions are supported 
by  precise numerical solutions for the same model, 
which demonstrate the very good performance 
of the mean-field CPA approach,  both for the density of 
states and the spectral density in a plane-wave basis 
(dynamical structure factor) in the 
whole energy range,  except for the high-frequency tail region 
containing localized states. 

The rest of the paper is arranged in the following 
manner. 
The model is defined in Sec.~\ref{s1}. 
The properties of the effective 
mean field are analysed in 
Sec.~\ref{s22}. 
The spectral densities and the scattering properties 
of the external plane waves  are discussed 
in Sec.~\ref{s3}. 
The conclusions are given in Sec.~\ref{s4}. 

%
%
%
%
\section{Model}
\label{s1}
\subsection{Hamiltonian}
\label{s1_1}

Atomic vibrations in the harmonic approximation 
can be treated in the classical limit  \cite{Elliott_74}. 
Thus the problem is reduced to the Hamiltonian formalism, 
in which the dynamical operator plays the  role of the Hamiltonian,  
and the  squared frequency, $\omega^2$, is an energy, 
$\varepsilon =  \omega^2$. 
The  relevant Hamiltonian 
(dynamical operator) for vector vibrations 
of $N$ interacting atoms of unit masses ($m_i=1$) 
can be written in the following form: 
\begin{equation}
{\hat \H}=\sum\limits_{i\alpha,j\alpha'} H_{i\alpha,j\alpha'}
|i,\alpha\rangle \langle j,\alpha' |
\ ,  
\label{e1_3}
\end{equation}
where we have used  the site orthonormal  basis of  
vectors $|i,\alpha \rangle$,  the only non-zero component 
of which is related to the displacement of atom $i$ 
($i=1, \dots, N$) 
along the Cartesian direction $\alpha$ ($\alpha=1, \dots, D$).  
The (dynamical) matrix elements $ H_{i\alpha,j\alpha'}$,  
being the second partial derivatives of the interatomic 
 potential,  obey the following sum rules 
\cite{Maradudin_71}: 
\begin{equation}
H_{i\alpha,i\alpha'}
= -  
\sum\limits_{j\ne i} H_{i\alpha,j\alpha'} 
\ .  
\label{e1_7}
\end{equation}
These  sum-rule correlations between the 
elements of the $D\times D$  diagonal and off-diagonal 
blocks 
distinguish  the  vibrational 
problems from   the electron problems described by  
similar Hamiltonians 
\cite{Economou_83:book,Gonis_92:book} 
(see Sec.~\ref{s3_e}).   
The matrix elements  
 of the Hamiltonian (\ref{e1_3}), generally speaking, 
are  functions of all equilibrium atomic positional vectors,  
$\r^{(0)}_i$. 

Our aim is to solve the eigenproblem for the 
Hamiltonian (\ref{e1_3}). 
This can be easily done for a crystal 
\cite{Maradudin_71} but is not so simple 
for a disordered structure. 
The analysis starts with a choice of the vibrational  
model for a disordered system. 
This could be:  (i) a crystalline lattice of atoms 
of random masses (mass disorder) and/or
 connected by random springs (force-constant disorder) 
\cite{Schirmacher_98,Martin_Mayor_00,Taraskin_01:PRL,Kantelhardt_01}; 
(ii) positionally disordered structures created, for example,  
by random atomic displacements around crystalline sites 
\cite{Martin_Mayor_01:JCP,Grigera_01,Grigera_01:BP}, or 
by other ways, e.g. the Bethe lattice or by bond switching 
\cite{Barkema_00};  
(iii) atomic structures, created e.g. by molecular dynamics (classical 
or {\it ab-initio}), in which intrinsic positional disorder defines 
the force-constant disorder (see e.g. 
Refs.~\cite{Sarnthein_95,Cappelletti_95,Nakhmanson_00,Vollmayr_96,Taraskin_97:PRB2}). 
The simplest models are from the first class. 
Analytical approaches can be easily developed for them 
due to the existence of a well-defined reference crystalline system 
(i.e. the same lattice without disorder) 
\cite{Elliott_74,Ehrenreich_76,Economou_83:book,Gonis_92:book}. 
Surprisingly at first sight (see below), 
 lattice models with force-constant 
disorder mimic very well the main features 
of topologically disordered glasses and even 
liquids,  e.g. the boson peak, localized band tails, 
main peaks in the density of states and 
zero-energy singularity in the instantaneous spectrum  
\cite{Schirmacher_98,Martin_Mayor_00,Taraskin_01:PRL,Simdyankin_00,
Taraskin_02:PRB_sing}. 
The known drawback of lattice models
 is related to the $k^4$-frequency dependence 
of the width of the dynamical structure factor 
\cite{Sheng_94,Schirmacher_98,Martin_Mayor_00,Grigera_01} 
as compared to the parabolic behaviour found 
experimentally (see e.g. \cite{Foret_96}). 
This $k^4$-dependence leads to a parabolic 
frequency scaling of the dynamic structure factor, 
$S(k,\omega) \propto \omega^2$ for 
$\omega \to 0$, which is in contrast to the 
frequency-independent $S(k,\omega)$ found 
for  topologically disordered models 
\cite{Gotze_00,Ruocco_00}. 

\subsection{Distribution of force constants}
\label{s1_2}

In this paper, we consider as a vibrational model 
 the crystalline f.c.c. lattice of atoms, in which 
the nearest neighbours are  connected 
by  springs that are unstretched in equilibrium. 
The matrix elements of the Hamiltonian (\ref{e1_3}) 
in this case have the following form: 
\begin{equation}
H_{i\alpha,j\alpha'} =
- \kappa_{ij}  ( {\hat \r}_{ij}^{(0)})_\alpha 
({\hat\r}_{ij}^{(0)})_{\alpha'} \ , \ \ \ \ \ 
\mbox{if} \ \ \ 
i\ne j 
 \ ,  
\label{e1_6}
\end{equation}
where 
$ {\hat \r}_{ij}^{(0)} = 
(\r_j^{(0)}  -\r_i^{(0)})/ |\r_j^{(0)}  -\r_i^{(0)}|   $  
is the unit connection vector 
between atoms $i$ and $j$ 
in equilibrium. 
The spring constants, $\kappa_{ij}$, are supposed 
to be independent of atomic positions and are random 
values taken from a certain probability 
distribution, $\rho(\kappa)$. 
Different functional forms of $\rho(\kappa)$ 
have been used for various problems, e.g. 
a combination of two $\delta$-functions 
\cite{Webman_81,Odagaki_81,Schirmacher_99,Martin_Mayor_00}, 
 (also used  
in the 
rigidity-percolation problem \cite{Feng_85}) and 
the normal (gaussian) 
\cite{Schirmacher_98}, box 
\cite{Taraskin_01:PRL} and hyperbolic 
\cite{Kantelhardt_01} distributions  for glassy problems. 
Below we use the box distribution: 
\begin{equation}
\rho(\kappa)=\frac{1}{2\Delta}
\left[
\theta(\kappa-\kappa_0+\Delta) - 
\theta(\kappa-\kappa_0- \Delta) 
\right] 
\ ,
\label{e1_2}
\end{equation}
where the Heavyside function $\theta(x<0)=0$ and 
$\theta(x\ge 0)=1$, $\kappa_0$ is the spring 
constant  in the reference crystal  
 and $2\Delta$ 
is the full width of the  distribution. 
The particular form  of the peak-shaped distribution 
 is not important for all results 
presented below from a qualitative point of view, 
 as we have checked 
by using the normal distribution instead of the box one. 
The box distribution also has the  advantage of 
not needing a lower cut-off to ensure positive values of force 
constants, as in the case for the normal and hyperbolic distributions. 

If the distribution given by Eq.~(\ref{e1_2}) is wide  enough, i.e. 
$\Delta > \kappa_0$ ($\kappa_0=1$ below), then  negative 
spring constants appear in the system which could result in 
 vibrational instability of the model in equilibrium, 
i.e. the occurrence of negative eigenvalues. 
Such a situation is of direct relevance 
\cite{Taraskin_02:PRB_sing} to 
an instantaneous mode analysis in liquids 
\cite{Wu_92,Wan_94,Keyes_97:review}.  

The vibrational spectrum of a disordered system is 
defined by the properties 
of the Hamiltonian (dynamical) matrix. 
In the case of a crystal, the Hamiltonian matrix is 
well structured (e.g. cyclic band diagonal). 
For topologically disordered glasses 
with conserved local order, the Hamiltonian 
matrix is structured similarly to that 
for the crystalline counterpart (with the same 
local order) but now, due to the presence of topological 
disorder: 
(i) the positions of elements are 
distributed about their 
positions in the crystalline matrix (locational disorder) and 
(ii)  the values of the elements are also 
distributed because of different interatomic distances. 
The first effect is completely ignored in the 
disordered-lattice models and the total influence of disorder 
is attributed entirely to the second effect. 
The question is, how good is this approximation?  

One criterion  relates to similarities  
in  the probability distributions for the values 
of the matrix elements for both types of models. 
We have compared the distributions 
of the Hamiltonian matrix elements for two vibrational 
models: (i)  the f.c.c. disordered-lattice model 
discussed above and described 
by the matrix elements 
$H_{i\alpha,j\alpha'}^{\rm lat}$ obeying 
Eq.~(\ref{e1_6}) and (ii) a topologically disordered 
model of a single-component glass with 
predominantly icosahedral order (IC-glass) 
obtained by classical molecular dynamics 
(see Refs.~\cite{Simdyankin_00,Simdyankin_01} for 
more detail). 
In the latter model, the Hamiltonian matrix 
elements, 
$H_{i\alpha,j\alpha'}^{\rm glass}$, are defined 
by the pairwise interatomic potential \cite{Dzugutov_92:glass}, 
$V(|\r_{ij}^{(0)}|)$, its spatial derivatives, 
$V'(|\r_{ij}^{(0)}|)$ and $V''(|\r_{ij}^{(0)}|)$,
and the  relative atomic 
positions, 
\begin{equation}
H_{i\alpha,j\alpha'}^{\rm glass} = 
- ({\hat \r}_{ij}^{(0)})_\alpha  
  ({\hat \r}_{ij}^{(0)})_{\alpha'} 
\left[ 
V''( |\r_{ij}^{(0)}|)
 - 
\frac{V'(|\r_{ij}^{(0)}|)}{|\r_{ij}^{(0)}|}
\right] -  
\delta_{\alpha,\alpha'} 
\frac{V'( |\r_{ij}^{(0)}|)}{|\r_{ij}^{(0)}|}
\ . 
\label{Hij_glass}
\end{equation}

The Hamiltonian matrix contains $D\times D$ diagonal 
and off-diagonal blocks, the diagonal 
($H_{i\alpha,i\alpha}$ and 
$H_{i\alpha,j\alpha}$) and off-diagonal 
($H_{i\alpha,i\alpha'}$ and 
$H_{i\alpha,j\alpha'}$) elements being distinctively 
distributed. 
For the diagonal elements in the off-diagonal blocks, 
it is convenient to compare the traces taken with 
opposite signs. 
They represent the force constants, $ 
\kappa_{ij}^{\rm lat}= 
\sum_\alpha H_{i\alpha,j\alpha}^{\rm lat} 
$ 
and $ 
\kappa_{ij}^{\rm glass}=
\sum_\alpha H_{i\alpha,j\alpha}^{\rm glass} = 
 (V''( |\r_{ij}^{(0)}|) + 
2V'( |\r_{ij}^{(0)}|)/|\r_{ij}^{(0)}|)$,   
for interactions 
beween atoms $i$ and $j$ in lattice and glassy models, respectively.  
Similarly the traces of diagonal blocks correspond 
to the total force constants, 
$\kappa_{ii}^{\text{lat}}$ 
and $\kappa_{ii}^{\text{glass}}$,  
for atom $i$  in lattice and glassy models, respectively,  where 
$
\kappa_{ii}^{\text{lat(glass)}}=  
\sum_j \kappa_{ij}^{\text{lat(glass)}}
$.  

\begin{figure}[b!] 
\centerline{\epsfig{file=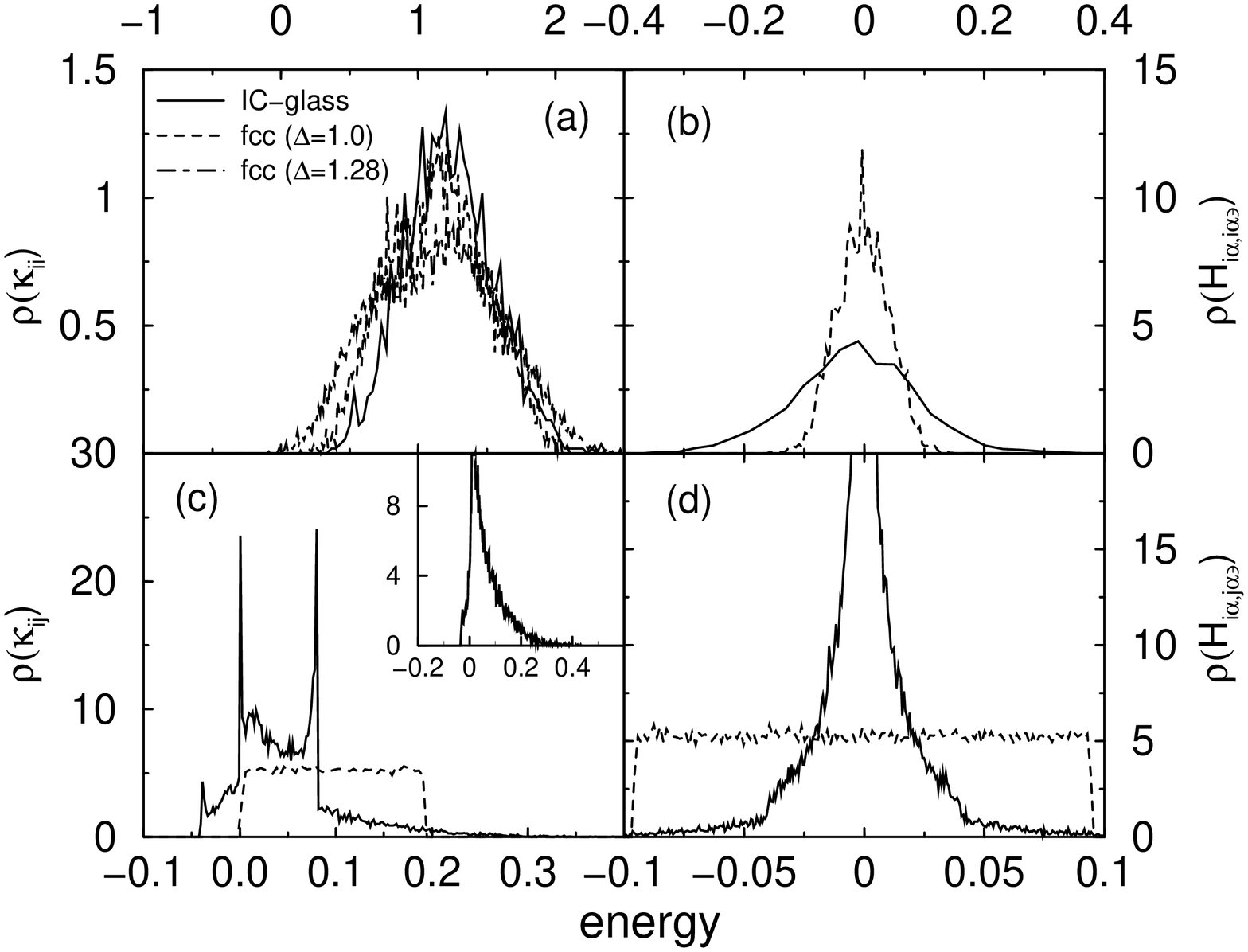,width=8truecm,angle=270}}
\vskip 20pt
\caption[]{The distribution of the 
matrix elements (scaled by the spectral band widths) 
 of the vibrational Hamiltonian matrix 
for the IC-glass (solid line) and disordered 
f.c.c. lattice 
(dashed ($\Delta =1$) and dot-dashed ($\Delta =1.28$) lines):  
(a) 
$\rho(\kappa_{ii})$ for 
the traces of  the diagonal blocks; 
(b) 
$\rho(H_{i\alpha,i\alpha'})$ for 
the off-diagonal elements in the diagonal blocks; 
(c) 
$\rho(\kappa_{ij})$ for 
the traces taken with opposite sign  in the off-diagonal blocks; 
(d) 
$\rho(H_{i\alpha,j\alpha'})$ for 
the off-diagonal elements in the off-diagonal blocks. 
The inset in (c) shows $\rho(H_{i,j}^{\rm glass})$ for nearest-neighbour 
interactions only.  
} 
\label{distr}
\end{figure}

The four distributions for both models
presented in Fig.~\ref{distr}(a)-(d)  show 
qualitatively the same features. 
The distributions of the diagonal elements 
in the diagonal blocks, 
$\rho(\kappa_{ii }^{\rm glass})$ and 
$\rho(\kappa_{ii }^{\rm lat})$, are both
peak-shaped with  
remarkably comparable peak positions and  peak widths
(see Fig.~\ref{distr}(a)). 
The distributions of the off-diagonal elements in the 
diagonal blocks are both centred around zero 
and have comparable widths 
(see Fig.~\ref{distr}(b)). 
The distributions of the elements for the off-diagonal 
blocks are much more model dependent. 
The solid line in Fig.~\ref{distr}(c) represents 
the distribution $\rho(\kappa_{ij}^{\rm glass})$ for  all 
interacting atomic pairs. 
The distribution function has a complicated form,  
 the features of which can be  understood 
in terms of the functional forms of the interatomic 
potential 
and the atomic  pair correlation function. 
For example, if  nearest-neighbour interactions 
only are taken into account 
(as in the disordered-lattice model), then the 
distribution function $\rho(H_{ij}^{\rm glass})$ has 
a single peak 
(see the inset in Fig.~\ref{distr}(c)), which corresponds 
to the first peak in the pair correlation function 
\cite{Dzugutov_92:glass}. 
The simplest model distribution which can mimic the 
function  $\rho(\kappa_{ij}^{\rm glass})$ 
(at least its scale)  is the 
box distribution (the dashed line in 
Fig.~\ref{distr}(c)).  
The distribution of the off-diagonal elements in the 
off-diagonal blocks 
(see Fig.~\ref{distr}(d))
has a maximum at zero and is symmetric about it. 
The disordered-lattice models are intrinsically anisotropic, 
so that the distribution function, 
$\rho(H_{i\alpha,j\alpha'}^{\rm lat})$, can look  
quite different to  
$\rho(H_{i\alpha,j\alpha'}^{\rm glass})$. 
For example, for  the ordered  f.c.c. lattice 
($\Delta=0$), it consists of  two $\delta$-functions 
symmetrically located around zero. 
With increasing disorder, these $\delta$-functions 
are transformed into two  box distributions 
which merge at $\Delta=1$, forming the single 
box distribution shown in  Fig.~\ref{distr}(d) by 
the dashed line. 
For larger $\Delta > 1$, 
these two boxes overlap around zero, thus resulting 
in a box-shaped peak in this region, i.e. the 
distribution function $\rho(H_{i\alpha,j\alpha'}^{\rm lat})$ 
becomes closer in shape to the peak-shaped function 
$\rho(H_{i\alpha,j\alpha'}^{\rm glass})$. 

The above comparison 
demonstrates  that the very simple 
disordered-lattice model with a box 
distribution of force constants is able to reproduce 
correctly the ranges of the distributions 
of the Hamiltonian matrix elements for a 
topologically disordered glass with  minimal short-range order 
(i.e. the IC-glass). 
This leads  to the plausible conjecture that 
the physical origins of  different  phenomena, such as 
the boson peak, Ioffe-Regel crossover and secondary 
peak in the dynamic structure factor (analysed below), 
localization-delocalization 
transition, etc.,  are  common for topologically 
disordered glasses and disordered lattices.

The other important point to emerge concerns 
the possibility of estimating the degree of disorder 
in disordered-lattice models at which they are able 
to mimic  glassy models. 
For small disorder,  all the distributions for 
the disordered-lattice model presented 
in Fig.~\ref{distr} are quite narrow and are not 
comparable in width with the distributions for the glassy model. 
They become similar only for sufficiently large 
disorder ($\Delta \sim 1$). 

Below we use the simplest (box) distribution  
of  force constants which allows us to analyse 
the problem analytically and which 
mimics some of the features of topologically 
disordered glasses by disordered lattices. 
A natural extension of the lattice models is in 
the use of more realistic force-constant distributions 
for crystalline counterparts of topologically disordered 
glasses and a comparison of the resulting properties of such 
models with the corresponding glasses. 
A possible way to mimic  disorder in glasses is 
to introduce  thermal (positional) disorder 
in the crystalline counterpart. 
We have done this for the IC-glass 
and its crystalline counterpart, the Frank-Kasper $\sigma$-phase 
\cite{Simdyankin_02:PRB}. 
A $\sigma$-phase lattice has been heated 
by means of molecular dynamics and the 
dynamical matrix for  the instantaneous cofiguration 
has been calculated. 
The elements of such a matrix do not exhibit 
locational disorder and only their values are distributed. 
Using temperature as a control parameter it is 
possible to obtain  the distributions of matrix elements  very 
similar to those ones for the  topologically disordered 
IC-glass (see Fig.~6 in Ref.~\cite{Simdyankin_02:PRB}).  
The vibrational density of states (VDOS) for instantaneous 
configurations of the $\sigma$-phase, 
as a consequence of such a similarity, resembles 
 the VDOS 
for the IC-glass \cite{Simdyankin_02:PRB,Simdyankin_00}. 
This supports the use of lattice models for mimicking 
topologically disordered structures. 
We should mention, however, that  
the  elements of the dynamical matrix 
(belonging to the Euclidian ensemble \cite{Mezard_99}) 
for positionally disordered crystalline counterparts 
are correlated with each other within a certain row. 
These correlations are absent for the 
simple box distributions used below. 

\subsection{Coherent potential approximation}
\label{s1_3}

\begin{figure}[b!] 
\centerline{\epsfig{file=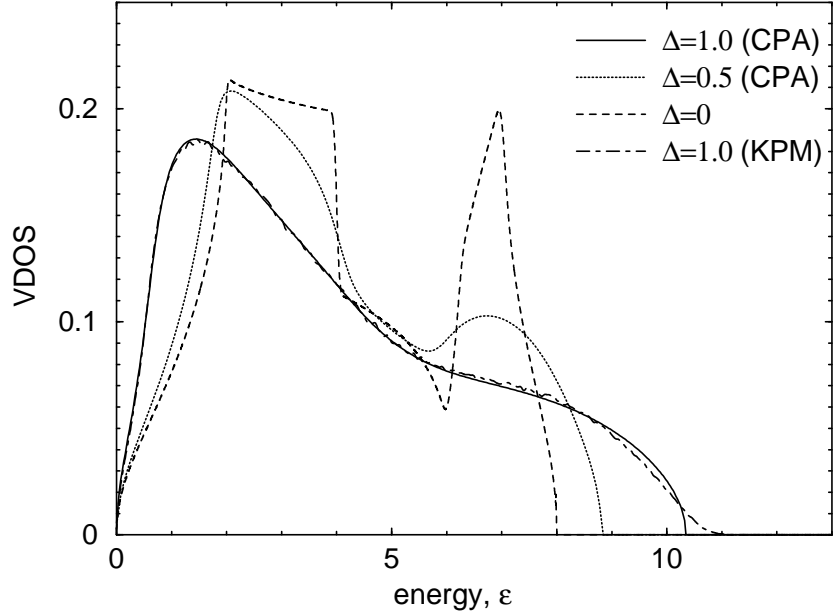,width=8truecm,angle=270}}
\vskip 20pt
\caption[]{
The vibrational density of states (VDOS) versus energy 
in the crystalline (dashed line) f.c.c. lattice and in 
f.c.c. lattices with  force-constant disorder characterized 
by different widths (as marked), $\Delta$, of the box distribution. 
The curves for the disordered models have been obtained 
by the CPA and the kernel 
polynomial method (KPM) \protect\cite{Silver_97}, as described in the text. 
} 
\label{VDOS}
\end{figure}

The solution of the eigenproblem for the 
Hamiltonian (\ref{e1_3}) is known for a crystal, 
when $\Delta = 0$ in Eq.~(\ref{e1_2}). 
The eigenmodes then are plane waves 
$|\k,\beta\rangle$, characterized by the wavevector $\k$, 
 the branch number $\beta$ and the eigenenergies 
$\varepsilon_{\k\beta}$. 
The VDOS, 
$ 
g^{\rm cryst}(\varepsilon) = 
(DN)^{-1} \sum_{\k,\beta} 
\delta(\varepsilon - \varepsilon_{\k\beta}) 
$,  
 forms a band (see the dashed line in 
Fig.~\ref{VDOS}) containing 
 van Hove singularities,  
and behaves as $ (\varepsilon)^{(D/2)-1}$ 
around the band edges \cite{Maradudin_71}.  
 Disorder ($\Delta > 0 $) brings  new features to  
the spectrum (see Fig.~\ref{VDOS}): 
(i) the van Hove singularities are washed out;  
(ii) a high-frequency tail containing localized states occurs;  
(iii) extra states appear  in the low-energy regime.

The exact solution of the vibrational 
problem with random spring 
constants is not known, 
but a number of  approximate methods, 
both for electrons and vibrations,  
have been developed 
\cite{Elliott_74,Economou_83:book,Efetov_97:book,Mirlin_00:review}. 
We are interested in the global features of the spectrum,  except 
the tail region  where the states are localized. 
An  adequate approach in this case   is a mean-field  treatment 
within the coherent potential approximation (CPA) 
\cite{Elliott_74,Ehrenreich_76,Economou_83:book},  which we use below. 
The CPA is known to be very successful  in a description 
of the electronic and vibrational  
properties of substitutional alloys 
and is expected to work very well in the whole 
energy range, except the far tails containing 
strongly localized states, where field-theoretical 
approaches are much better \cite{Mirlin_00:review}.

The Hamiltonian (\ref{e1_3}) can be rewritten  in the 
bond representation, 
\begin{equation}
{\hat \H}=\frac{1}{2}\sum\limits_{i,j} 
\kappa_{ij} |ij\rangle\langle ij| 
\ , 
\label{e1_8}
\end{equation}
with 
$
|ij\rangle = \sum_\alpha  
({\hat \r}_{ij}^{(0)})_\alpha 
(|i,\alpha\rangle - |j,\alpha\rangle) 
$, 
which is a convenient form for a subsequent 
mean-field treatment. 
The VDOS of disordered systems, $g(\varepsilon)$,  
\begin{equation}
g(\varepsilon)=\frac{1}{ND}{\rm Tr}{\hat \A}(\varepsilon)
= \frac{1}{ND}
\sum_{\k\beta}
\left\langle  \sum_d  
|\langle d| \k,\beta\rangle|^2
\delta(\varepsilon -\varepsilon_d) 
\right\rangle
\ ,   
\label{e2_7}
\end{equation}
can be defined 
via the spectral-density operator 
\cite{Ehrenreich_76}, 
${\hat\A}(\varepsilon)={\big\langle} 
\delta (\varepsilon-{\hat \H}) {\big\rangle}
$, 
where  the angular  brackets, 
$\langle \dots \rangle$, 
denote averaging over random spring constants. 
The coefficients,  $  |\langle d| \k,\beta\rangle|^2 $, in 
Eq.~(\ref{e2_7}) have a simple physical meaning  
\cite{Ehrenreich_76}; 
they define the weight of the disordered  eigenstate $|d\rangle$ 
in the crystalline state $|\k,\beta\rangle$, which is a 
plane wave of a certain polarization. 
The importance  of the spectral-density operator 
in the plane-wave  basis is related to the fact that  the matrix elements 
${\hat \A}_{\k\beta}(\varepsilon) = 
\langle\k\beta|{\hat\A}|\k\beta\rangle $ are  proportional 
to the dynamical structure factor  
\cite{Martin_Mayor_00} (see Sec.~\ref{s3_dd})  which can be 
measured experimentally \cite{Elliott_90:book}. 

Both the VDOS, $g(\varepsilon)$, 
 and the spectral density,  
${\hat \A}_{\k\beta}(\varepsilon)$,  
can be found for vector vibrations within the 
mean-field approximation (see Appendix A): 
\begin{equation}
 g(\varepsilon) = 
-\frac{1}{\pi}\mbox{Im}\left[\frac{1}{z(\varepsilon)}G^{\rm cryst}
\left(\frac{\varepsilon}{z(\varepsilon)}\right)\right] 
\ , 
\label{e2_17} 
\end{equation}
and 
\begin{equation}
 A_{\k\beta}(\varepsilon) = 
-\frac{1}{\pi} 
\frac{z''(\varepsilon)\varepsilon_{\k\beta}}
{
\left[\varepsilon - z'(\varepsilon)\varepsilon_{\k\beta}
\right]^2 + 
\left[
z''(\varepsilon)\varepsilon_{\k\beta}
\right]^2
}
\ , 
\label{e2_18} 
\end{equation}
via the crystalline Green's function, 
$G^{\rm cryst}$,  and the complex effective 
mean field, $z(\varepsilon)=z'(\varepsilon)+i z''(\varepsilon) 
= {\tilde\kappa}(\varepsilon)/\kappa_0$ 
(the dimensionless effective force constant),  
which can be found from the solution of the 
self-consistent Eq.~(\ref{e2_13}). 

%
%
%
%
\section{Effective field }
\label{s22}

As seen from Eqs.~(\ref{e2_17})-(\ref{e2_18}),  
the spectral properties of a disordered lattice, 
within the CPA,  are characterized by the 
energy dependence of the 
effective dimensionless force constant (effective field), 
$z(\varepsilon)$. 
In this section, we discuss the properties  of the 
effective field obtained by the solution of the 
self-consistent equation (\ref{e2_13}) (see 
Appendix A) for different 
values of disorder and in different energy ranges. 
The most important regime from the viewpoint of comparison 
with experimental data for the dynamical structure factor  
is the low-energy limit ($\varepsilon \to 0$),  analyzed  
below. 

For the  particular choice of the box-like 
force-constant probability 
distribution given by Eq.~(\ref{e1_2}), 
the integration in 
Eq.~(\ref{e2_13}) can be done analytically,  
resulting in the following expression: 
\begin{equation}
1+\frac{1}{2\Delta\alpha(\varepsilon)}
\mbox{ln}\left[\frac{
1+\Delta-z(\varepsilon)-\alpha^{-1}(\varepsilon)}
{1-\Delta-z(\varepsilon)-\alpha^{-1}(\varepsilon)} 
\right]
=0 \ ,
\label{e22_1} 
\end{equation}
with 
\begin{equation}
\alpha(\varepsilon)\equiv \kappa_0 
\langle ij|{\tilde {\hat\G}}|ij\rangle
\ .
\label{e22_1aa} 
\end{equation}
Eq.~(\ref{e22_1}) should be solved with respect to the complex 
effective field $z(\varepsilon)$. 
Without disorder ($\Delta = 0$), the solution 
is trivial: $z'=1$ and $z''=0$. 
For finite disorder, the solution 
 can be found numerically in the general case and analytically in some 
limiting cases. 

%
%
%
\subsection{General case}
\label{s22a} 

The numerical solution of Eq.~(\ref{e22_1})  for the 
real and imaginary parts of the effective field 
for different values of disorder in the whole energy range 
 is presented in Fig.~\ref{Z_eff}. 
As seen from Fig.~\ref{Z_eff}(a), the real part of the effective 
spring constant  varies around its crystalline value 
$z'=1$, being less than unity  in the lower part 
of the energy band and greater than unity  in the upper part of the band. 
This is the expected behaviour, because the value of 
$z'$ describes the level-repelling effect for the 
bare crystalline states when disorder is introduced in the 
system \cite{Taraskin_01:PRL} 
(see Sec.~\ref{s3_1a}). 
The deviation of $z'$  from unity increases with 
increasing disorder, which reflects the more pronounced 
degree of level repelling in more highly disordered systems. 

The imaginary part of the effective field 
is non-zero and negative only in the band region. 
It basically reproduces the shape of the 
VDOS  (Eq.~(\ref{e2_17})) and describes the widths of the 
spectral-density peak  
(Eq.~(\ref{e2_18}))
which is proportional  to the 
inverse life-time of plane waves in disordered 
structures (see Sec.~\ref{s3_b}). 
The magnitude of $z''$ increases with increasing 
disorder,  thus indicating the shortening of the 
plane-wave life-time in strongly disordered 
lattices. 
The imaginary part of the effective field approaches 
zero in the low-energy limit, 
$z''(\varepsilon \to 0) \to 0$,   only if the disorder 
is low enough, for $\Delta \le \Delta_*$, when 
the structure in equilibrium is mechanically stable 
within the CPA (i.e. there are no negative eigenvalues).

%
%
%
\subsection{Low-energy limit}
\label{s22b} 

Using the energy dependence of the effective field 
in the low-energy regime (see Appendix B), 
the expression (\ref{e2_17}) 
for the VDOS can be rewritten as 
\begin{equation}
 g(\varepsilon) = 
\frac{1}{z'(\varepsilon)}g^{\rm cryst}
\left(\frac{\varepsilon}{z'(\varepsilon)}\right)  
\simeq  
 \frac{1}{\left(z'(0)\right)^{D/2}}
g^{\rm cryst}(\varepsilon) 
\simeq   \frac{\chi_{\rm Deb}}{\left(z'(0)\right)^{D/2}} 
\varepsilon^{(D/2)-1} 
\ . 
\label{e22_7} 
\end{equation}
As follows from Eq.~(\ref{e22_7}), the 
low-energy VDOS in  systems with lattice disorder 
functionally 
behaves the same as in the reference crystal, i.e. 
$ g^{\rm cryst}(\varepsilon) \propto 
\varepsilon^{(D/2)-1}$ (see Fig.~\ref{VDOS_log}), 
and differs from 
the crystalline VDOS just by the factor $(z'(0) )^{-D/2}>1$. 
In other words,  extra states, in addition to the 
crystalline ones, appear in the low-energy regime 
due to disorder-induced level-repelling effects 
\cite{Ehrenreich_76,Taraskin_99:JPCM,
Taraskin_99:PRB,Taraskin_01:PRL,Finkemeier_01,Simdyankin_02:PRB}.  
The relative density of these extra states is: 
\begin{equation}
 \frac{g(\varepsilon) - g^{\rm cryst}(\varepsilon) }
{ g^{\rm cryst}(\varepsilon)} \simeq 
\left[z'(0) \right]^{-3/2} - 1 
\  .   
\label{e22_8} 
\end{equation}
With increasing energy, the disordered VDOS 
increases in energy  
faster than $ \varepsilon^{(D/2)-1}$ and this results 
in the  boson peak, the origin of which has been 
discussed in Ref.~\cite{Taraskin_01:PRL}.

In the low-disorder limit (see Appendix B), 
$\Delta \to 0$, the  extra density of states 
(see Eq.~(\ref{e22_8})) in the low-energy range 
scales quadratically with disorder, 
\begin{equation}
 \frac{g(\varepsilon) - g^{\rm cryst}(\varepsilon) }
{ g^{\rm cryst}(\varepsilon)} \simeq 
\frac{D}{Z}\Delta^2 \ , \ \ \ \mbox{for}\ \ \ \Delta\ll 1 
\ ,  
\label{e22_11} 
\end{equation}
which has been  confirmed numerically 
(see Fig.~\ref{VDOS_0_vs_Delta}). 

%
%
%
\subsection{Critical disorder}
\label{s22d} 
All the derivations presented above 
are valid only for sufficiently small disorder, 
$\Delta <\Delta_*$. 
This restriction comes from Eq.~(\ref{e22_5}) 
(see Appendix B), 
which has solutions only for 
$\Delta \le \Delta_*$ (see Fig.~\ref{Delta_critical}). 
The critical value of disorder can be found numerically,  
giving  $\Delta_* \simeq 1.296$. 
The corresponding value of the real part of the 
effective field at criticality, 
$z'_*=z'(\Delta_*)$, can be found analytically 
using the fact that $(df/dz')_{\Delta=\Delta_*}=0$ (
see Eq.~(\ref{e22_5})), so that 
\begin{equation}
z'_* = 1 - \sqrt{1-\frac{1-\Delta_*^2}{\gamma}}
\ .  
\label{e22_12} 
\end{equation}
Eq.~(\ref{e22_12}) gives $z'_*\simeq 0.43$ for the 
f.c.c. lattice (with $\gamma \equiv (Z/2D)-1 = 1$ 
for $Z=12$ and $D=3$). 

When the disorder approaches the critical value where the VDOS behaves as a 
renormalized Debye function (see Eq.~(\ref{e22_7}))  
and the imaginary part of the effective force 
constant behaves as $z''\propto \varepsilon^{D/2}$ 
(see Eq.~(\ref{e22_6})), 
 the low-energy range  shrinks to zero 
(see Fig.~\ref{VDOS_log}). 
Above the critical disorder, both the 
imaginary part of the effective field 
and the effective VDOS are finite at zero energy, $z''(0)\ne 0$ 
and $g(\varepsilon = 0)\ne 0$, and negative eigenvalues 
appear  within the CPA. 
This means that the system becomes vibrationally unstable 
in equilibrium, even within the mean-field description. 
Such a situation corresponds to the spectrum 
of the dynamical matrix for  instantaneous metastable configurations 
in liquid  and glassy states. 
It turns out that the energy  spectrum of the instantaneous  
dynamical matrix in 
the unstable regime ($\Delta > \Delta_*$) exhibits 
a peculiar (singular) universal behaviour 
around zero energy,  which we have also 
found in numerical experiments on liquid and vitreous 
SiO$_2$ and the IC-liquid and IC-glass    
\cite{Taraskin_02:PRB_sing}. 

%
%
%
%
%
%
\section{Spectral density}
\label{s3}
%
%
%
%
%
\subsection{Shape of the spectral density}
\label{s3_1}
\begin{figure}[b!] 
\centerline{\epsfig{file=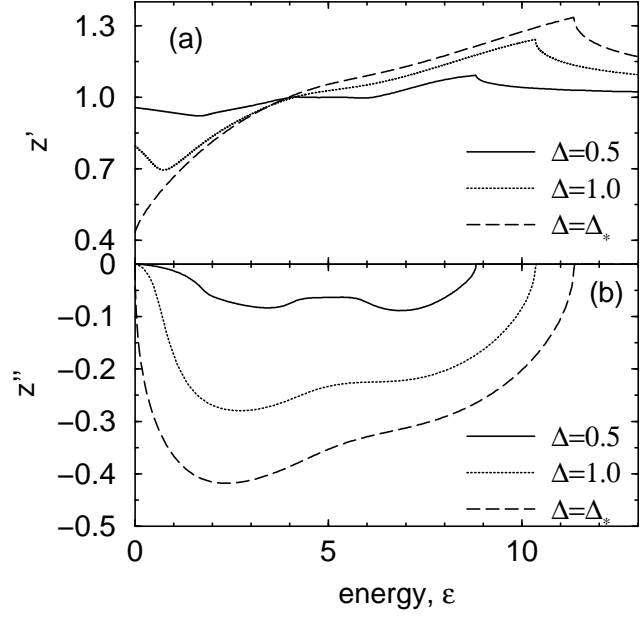,width=8truecm,angle=270}}
\vskip 20pt
\caption[]{
(a) The real, $z'$,  and (b) imaginary, $z''$, parts 
of the effective force 
constant versus energy for different values of disorder 
as marked.  
} 
\label{Z_eff}
\end{figure}

\begin{figure}[b!] 
\centerline{\epsfig{file=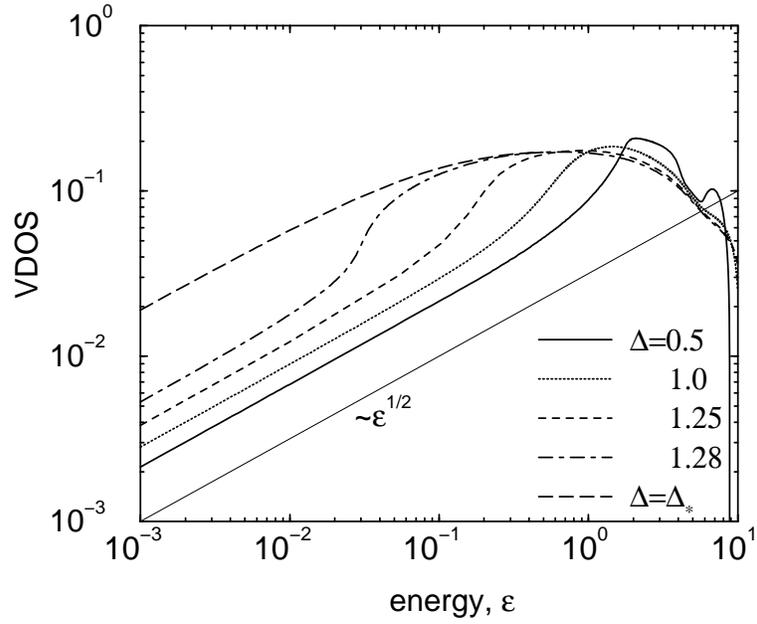,width=8truecm,angle=270}}
\vskip 20pt
\caption[]{
The VDOS, $g(\varepsilon)$, in the low-energy regime for different 
values of disorder as marked. 
The thin  solid line (guide for the eye) shows an 
$\varepsilon^{1/2}$ dependence. 
} 
\label{VDOS_log}
\end{figure}

\begin{figure}[b!] 
\centerline{\epsfig{file=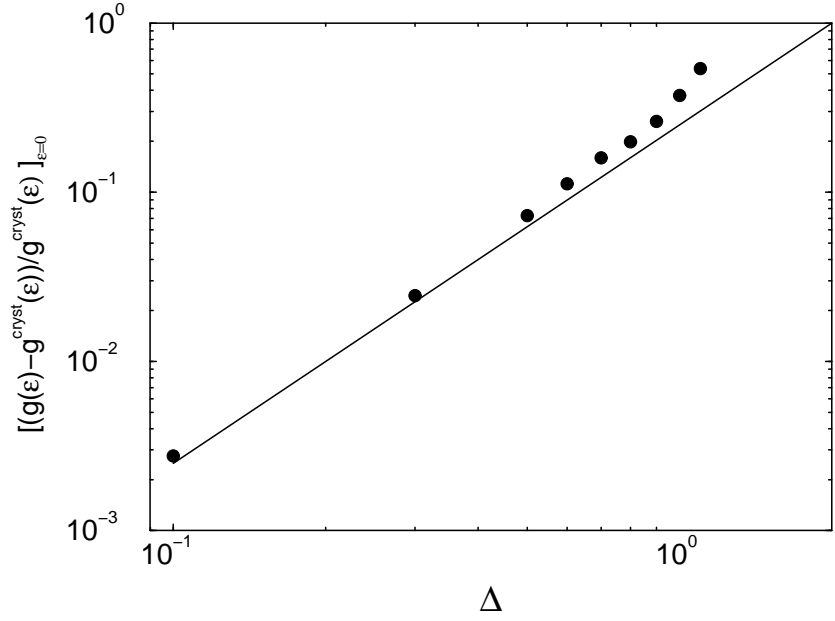, width=8truecm,angle=270}}
\vskip 20pt
\caption[]{
The relative density of extra states, 
$[g(\varepsilon)- g^{\rm cryst}(\varepsilon)]/ 
g^{\rm cryst}(\varepsilon) $, in the limit of  zero energy 
($\varepsilon \to 0$) 
versus disorder. 
The straight line indicates a quadratic dependence on $\Delta$. 
} 
\label{VDOS_0_vs_Delta}
\end{figure}

\begin{figure}[b!] 
\centerline{\epsfig{file=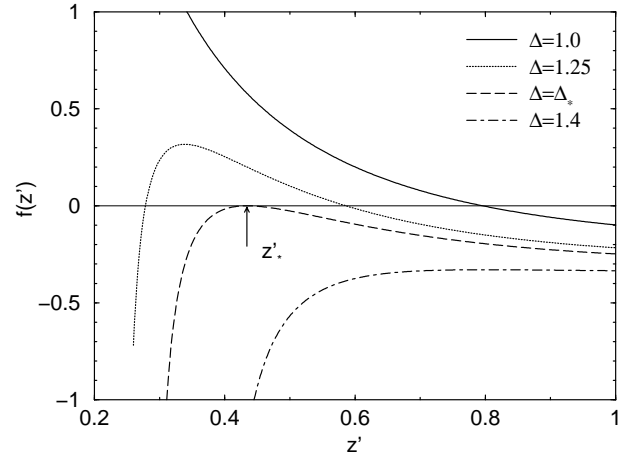,width=6truecm,angle=270}}
\vskip 20pt
\caption[]{
The evolution of the function $f(z')$ given by 
Eq.~(\protect\ref{e22_5}). 
The larger root of the equation $f(z')=0$
 existing only at $\Delta\le \Delta_*$ corresponds to the 
real part of the effective force constant. 
} 
\label{Delta_critical}
\end{figure}

The spectral density in the plane-wave representation, 
$A_{\k\beta}(\varepsilon)$,  
provides  information about the distribution 
of contributions of the different 
disordered eigenstates in a particular plane 
wave from the branch $\beta$ and with wavevector $\k$.  
In a crystal, the spectral density 
is obviously a $\delta$-function, 
$ 
A^{\rm cryst}_{\k\beta}(\varepsilon) = 
\delta(\varepsilon - \varepsilon_{\k\beta})
$. 
Disorder introduced in the system transforms 
$ A^{\rm cryst}_{\k\beta}(\varepsilon)$ into 
the disordered spectral density which has 
a shape given by Eq.~(\ref{e2_18}). 
The shape of  $ A_{\k\beta}(\varepsilon)$ 
depends on the energy behaviour of the 
effective field $z(\varepsilon)$. 
If we assume that the effective field 
is energy independent, then 
 the spectral density has the shape of a Lorentzian 
peak, the width of which is proportional 
to the imaginary part, $z''$,  of the effective field. 
Of course,  the effective field does depend on energy 
(see Fig.~\ref{Z_eff}), 
 but if the imaginary part of the effective 
field is small enough, which is true 
in the whole energy range for sufficiently 
small disorder,  and at least 
in the 
low-frequency regime ($\varepsilon_{\k\beta} \alt  
\varepsilon^{\rm IR}_{\k\beta} $, see below) 
at any disorder below the critical one ($\Delta < \Delta_*$), then 
the disordered spectral density still has the  shape 
of a narrow peak (see Fig.~\ref{SD_KPM}). 
The  position, $ \varepsilon^{\rm max}_{\k\beta} $, 
and full width at half-maximum, 
$\Gamma_{\k\beta}$, 
of such a  spectral-density peak 
can be estimated from the following equations: 
\begin{equation}
\varepsilon^{\rm max}_{\k\beta}\simeq z'(\varepsilon^{\rm max}_{\k\beta}) 
\varepsilon_{\k\beta} 
\ \ \ \mbox{and} \ \ \ 
\Gamma_{\k\beta}\simeq 2|z''(\varepsilon^{\rm max}_{\k\beta})| 
\varepsilon_{\k\beta} 
\ .     
\label{e3_1}
\end{equation}

 For a weak energy dependence of 
the real part of the effective field (see  Fig.~\ref{Z_eff}(a)), 
we can approximate Eq.~(\ref{e3_1})  giving: 
\begin{equation}
\varepsilon^{\rm max}_{\k\beta}
\simeq 
z'(\varepsilon_{\k\beta}) 
\varepsilon_{\k\beta}
\ .         
\label{e3_2}
\end{equation}
In the low-energy regime 
 ($\varepsilon_{\k\beta}\to 0$),   
the imaginary part 
of the effective field is $z''\propto 
\varepsilon^{D/2}$ according to 
Eq.~(\ref{e22_6}), so that 
\begin{equation}
\Gamma_{\k\beta} \propto  
\varepsilon_{\k\beta}^{(D/2)+1} 
\ .   
\label{e3_3} 
\end{equation}
This equation corresponds to the well-known 
Rayleigh law for the peak width, 
$ \Gamma_{\omega}(\omega_{\k\beta}) $, 
 in the $\omega$-representation 
\cite{Sheng_94,Schirmacher_98,Martin_Mayor_00}: 
\begin{equation}
\Gamma_{\omega}(\omega_{\k\beta}) = 
\frac{1}{2\omega} 
\Gamma_{\k\beta}(\varepsilon^2_{\k\beta})
\propto  
\omega_{\k\beta}^{D+1} 
\ ,   
\label{e3_3a} 
\end{equation}
or  $\Gamma_{\omega}(\omega_{\k\beta}) 
 \propto 
\omega_{\k\beta}^4$ in the three-dimensional  case 
(see Fig.~\ref{IR_new}). 

Eq.~(\ref{e3_2}) defines the dispersion law in 
disordered lattices, 
$ \omega_{\k\beta}^{\rm max} = 
\sqrt{\varepsilon_{\k\beta}^{\rm max}}  \simeq  
c_\beta^{\rm dis} k $ (for $\varepsilon_{\k\beta} \to 0$), 
which is basically the same as 
in the reference crystal except for the renormalized 
sound velocity, $ c_\beta^{\rm dis} \simeq 
\sqrt{z'(0)} c_\beta^{\rm cryst}$,  with 
$ c_\beta^{\rm cryst}$ being the sound velocity 
for branch $\beta$ in the reference crystal. 
Bearing in mind that $z'(0)<1$ (Fig.~\ref{Z_eff}(a)), it is evident that  
$c_\beta^{\rm dis} < c_\beta^{\rm cryst}$
 (see the inset in Fig.~\ref{IR_new}). 
We have found a similar relation for the model 
of the topologically disordered IC-glass,  where 
$c^{\rm dis}/ c^{\rm cryst} \simeq 0.72$ 
\cite{Simdyankin_02:PRB}. 
This value corresponds to $\Delta \simeq 1.28$ 
(see the inset in Fig.~\ref{IR_new}),  
for which the distribution of the diagonal elements 
in the diagonal blocks for the disordered-lattice model fits 
very well the similar distribution for the IC-glass 
(see Fig.~\ref{distr}(a)). 

%
%
%
%
%
\subsection{Level-repelling effects}
\label{s3_1a}

As seen from Eq.~(\ref{e3_2}), disorder results in a  
shift of the crystalline spectral density, characterized by the real part of the effective 
field, $z'$. 
The bare level moves downwards (upwards) if 
$z'<1$ ($z'>1$), 
i.e. when it is located in the lower (upper) 
 part of the band (see Fig.~\ref{Z_eff}(a)). 
The shift of the bare level, $\varepsilon_{\k\beta}$, by 
disorder is a consequence of 
level-repelling effects 
\cite{Ehrenreich_76,Taraskin_99:JPCM,Taraskin_99:PRB,Taraskin_01:PRL}. 
Indeed, disorder introduced in the system results 
in the appearance of non-zero interaction matrix elements 
between the bare states. 
These interactions lead to  standard quantum repelling 
between levels \cite{Landau_63,Maradudin_71}. 
If the interaction matrix elements are approximately 
the same for all levels, then the bare crystalline level, 
$\varepsilon_{\k\beta}$,  
from the lower part of the energy band  is 
shifted downwards just because the number of the states 
above the bare level (and repelling it downwards) is larger 
than the number of  states below it 
(and repelling it upwards). 
This explains qualitatively why the value 
of the real part of the effective field is $z'<1$ in the 
lower part of the band  and 
{\it vice versa}, $z'>1$ in the upper 
half of the energy band (see Fig.~\ref{Z_eff}(a)). 

We have proved quantitatively (analytically and numerically) 
 such a picture of  
level-repelling effects in the VDOS for 
the force-constant-disordered f.c.c. lattice 
\cite{Taraskin_01:PRL} and in 
the topologically disordered IC-glass  
\cite{Simdyankin_01}. 
A similar effect has also been found numerically 
in models of disordered Si \cite{Finkemeier_01}. 

\begin{figure}[b!] 
\centerline{\epsfig{file=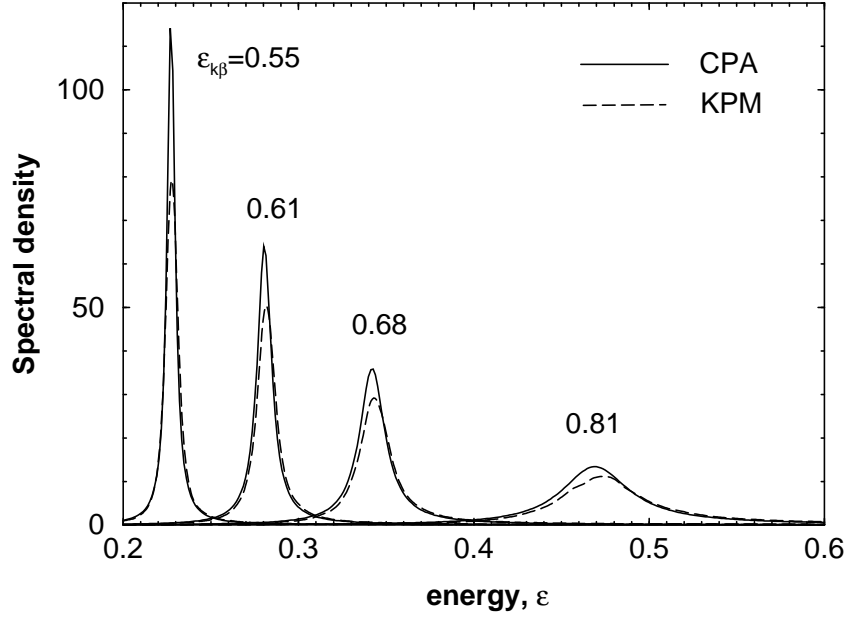,width=8truecm,angle=270}}
\vskip 20pt
\caption[]{
The spectral densities for vector vibrations 
in the f.c.c. lattice with force-constant disorder 
(the width of the box distributions $\Delta = 1$) 
calculated by the CPA (solid lines) and KPM (dashed lines). 
} 
\label{SD_KPM}
\end{figure}

\begin{figure}[b!] 
\centerline{\epsfig{file=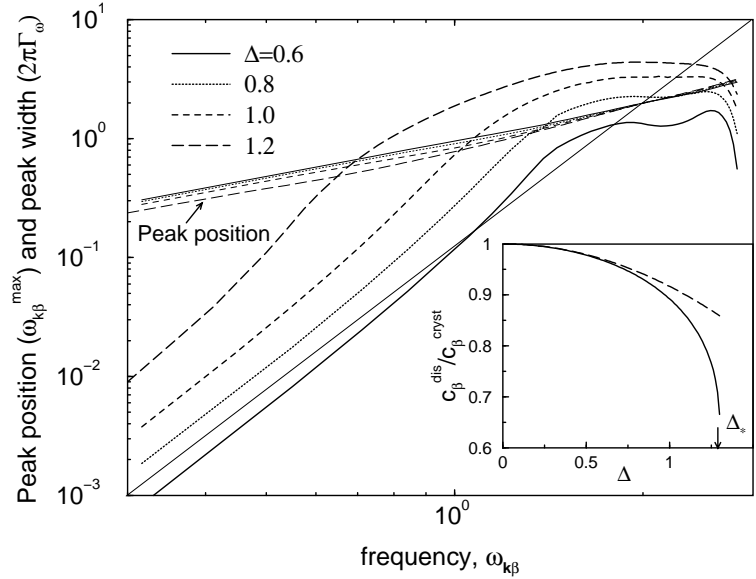,width=8truecm,angle=270}}
\vskip 20pt
\caption[]{
The peak positions, $\omega^{\rm max}_{\k\beta}$ (thin lines), 
and peak widths, 
$2\pi\Gamma_\omega$ (thick lines), versus 
plane-wave frequency, $\omega_{\k\beta}$, 
for different values of disorder as marked. 
The crossing points of the curves for peak positions 
and peak widths 
correspond to the Ioffe-Regel crossover frequency. 
The thin solid line (guide for the eye) 
shows an 
$\omega_{\k\beta}^4 $ dependence.  
The inset shows the ratio of the sound velocities  
in disordered and ordered lattices, 
$c^{\rm dis}_\beta/ c^{\rm cryst}_\beta$, 
versus degree of disorder (solid line). 
The dashed curve in the inset 
 corresponds to the low-disorder  approximation, 
$c^{\rm dis}_\beta/ c^{\rm cryst}_\beta \simeq 
1-D\Delta^2/3Z $, valid for $\Delta\ll 1$. 
The critical behaviour of  
$ c^{\rm dis}_\beta/ c^{\rm cryst}_\beta$ is evident 
for $\Delta \to \Delta_*$. 
}
\label{IR_new}
\end{figure}

An example of the spectral densities for 
different energies of external plane waves 
at relatively high disorder ($\Delta = 1$) 
is shown in Fig.~\ref{SD_KPM}. 
Their widths at sufficiently large $\varepsilon_{\k\beta}$  
become comparable to their positions and plane waves 
at such energies are very short-lived 
quasiparticles \cite{Taraskin_00:PRB_IR1,Taraskin_00:PRB_IR2}. 
The level-repelling effects are clearly seen 
in  pronounced shifts (to lower energies) of  the peak positions 
with respect to $\varepsilon_{\k\beta}$. 

We have also presented in Fig.~\ref{SD_KPM} the 
results of precise numerical solutions for the 
spectral densities obtained by the 
the kernel polynomial method (KPM) \cite{Silver_97}. 
The very good agreement between the CPA and 
KPM results supports the 
reliability of the mean-field treatment  
in obtaining  the spectral 
densities.

\subsection{Ioffe-Regel crossover and localization}
\label{s3_b}

Disorder
also broadens a bare level, so that the $\delta$-functions 
making up the crystalline VDOS 
are broadened into peaks 
 of finite width,  
 related to the imaginary part 
of the effective field  
(see Eq.~(\ref{e3_1})). 
This occurs because a plane wave $|\k,\beta\rangle$ is not 
an eigenstate in the disordered  lattice. 
So it decays with time and can be treated as a 
quasiparticle having  a finite life-time. 
This life-time is inversely proportional  to 
the width of the peak of the spectral 
density and this determines 
the physical meaning of  $z''(\varepsilon)$ 
\cite{Ehrenreich_76}. 

The scattering properties of  external plane 
waves in disordered lattices strongly 
depend on their bare energy. 
If the energy is low enough, 
the spectral-density peak is narrow  compared to its position, 
and the quasiparticles live for a long time,  
thus indicating a weak-scattering regime. 
For higher energies,  the width of the spectral density 
becomes comparable to the peak position (see Fig.~\ref{IR_new}).  
This means that the quasiparticles there 
decay quickly and 
the bare plane waves are in the region of strong 
scattering.   
The crossover between weak- and strong-scattering regimes 
is called the Ioffe-Regel crossover 
\cite{Ioffe_60,Taraskin_00:PRB_IR1,Taraskin_00:PRB_IR2}. 
The energy of the bare plane wave, 
$\varepsilon_{\k\beta}^{\rm IR}$, at which such a crossover 
occurs is the Ioffe-Regel crossover energy. 

The quantitative condition 
for the Ioffe-Regel crossover is  
\cite{Taraskin_00:PRB_IR2}, 
\begin{equation}
\frac{\omega^{\rm max}_{\k\beta}}{2\pi} 
\tau_{\k\beta} \sim 1 
\ ,
\label{e3_c_1}
\end{equation}
with $\omega^{\rm max}_{\k\beta} \equiv 
\sqrt{\varepsilon^{\rm max}_{\k\beta}}$ standing 
for the frequency  and 
$\tau_{\k\beta}$ being the life-time of the quasiparticle. 
The quasiparticle life-time can be easily evaluated 
in the weak-scattering regime from the 
time dependence of the   probability, 
$ |\langle \k,\beta;t|\k,\beta;0\rangle|^2 $, 
to find the system at time $t$ in the same plane-wave state  
$|\k,\beta\rangle$ as it was at $t=0$ 
\cite{Ehrenreich_76}, i.e.   
$
\langle|\langle \k,\beta;t|\k,\beta;0\rangle|^2
\rangle \simeq  
\cos^2(\omega^{\rm max}_{\k\beta} t) 
\exp(-t/\tau_{\k\beta})
$, 
where 
\begin{equation}
\tau_{\k\beta} 
= 
\frac{1}{|z''(\varepsilon_{\k\beta})|} \sqrt{
\frac{z'(\varepsilon_{\k\beta})}
{\varepsilon_{\k\beta}}
} \propto \varepsilon_{\k\beta}^{-(D+1)/2}
\ .   
\label{e3_c_6}
\end{equation}
The inverse life-time, $\tau_{\k\beta}^{-1}$, 
of course, coincides with the full width, 
$  \Gamma_\omega(\varepsilon_{\k\beta})  $, 
of the spectral density  
in the $\omega$-representation 
\cite{Taraskin_99:PRB}, 
$
\tau^{-1}_{\k\beta} 
= 
\Gamma_\omega(\varepsilon_{\k\beta}) 
\equiv 
\Gamma_{\k\beta}/(2\sqrt{\varepsilon^{\rm max}_{\k\beta}}) 
$, 
and the Ioffe-Regel criterion (\ref{e3_c_1}) 
can therefore be rewritten as 
\begin{equation}
\frac{ \omega^{\rm max}_{\k\beta}
(\varepsilon_{\k\beta})}{2\pi} 
\sim \Gamma_\omega( \varepsilon_{\k\beta} ) 
\ . 
\label{e3_c_8}
\end{equation}
This equation should be solved with respect to 
$\varepsilon_{\k\beta}$ (or 
$\omega_{\k\beta} = \sqrt{\varepsilon_{\k\beta}}$) 
and an estimate, 
$ \varepsilon_{\k\beta}^{\rm IR}$ (or 
$ \omega_{\k\beta}^{\rm IR}$), 
  for the 
Ioffe-Regel crossover energy (or frequency) can be found. 
Such a solution can exist at sufficiently 
large $ \omega_{\k\beta} $. 
Indeed, in the low-frequency limit, 
the width of the spectral density, 
$ \Gamma_\omega \propto \omega_{\k\beta}^{D+1}$,  
is much smaller than the peak position, 
$\omega^{\rm max}_{\k\beta} \propto 
\omega_{\k\beta}$, but $\Gamma_\omega$ grows 
quickly with increasing $\omega_{\k\beta}$ and can easily 
reach the peak-position value, 
if the disorder is large enough (see Fig.~\ref{IR_new}). 

\begin{figure}[b!] 
\centerline{\epsfig{file=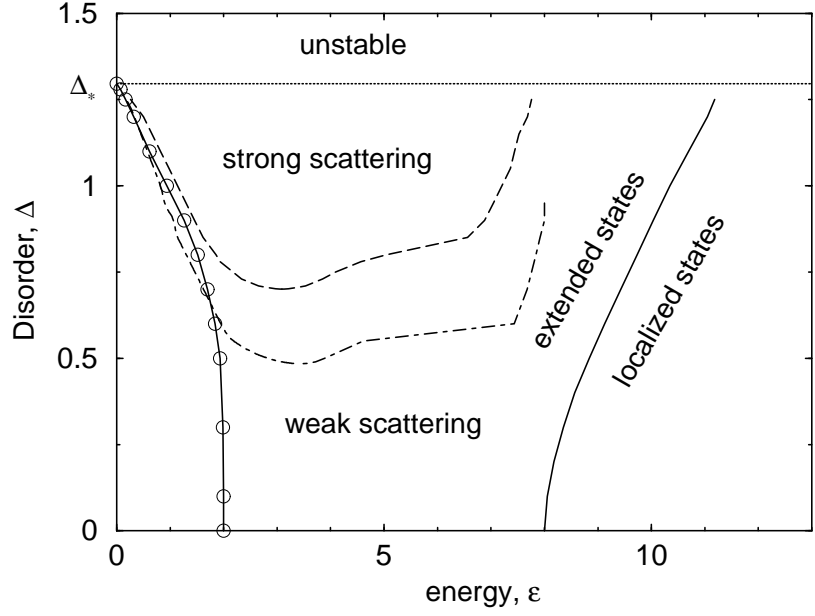,width=8truecm,angle=270}}
\vskip 20pt
\caption[]{
Phase diagram for vector vibrations in the f.c.c. 
lattice with force-constant disorder characterized by 
the box distribution. 
The dashed (dot-dashed) line corresponds to the 
Ioffe-Regel crossover plane-wave energy 
obtained from Eq.~(\protect\ref{e3_c_8}) (or from 
Eq.~(\protect\ref{e3_c_8}) in which the spectral-density 
peak width is decreased, say, by a factor $2$). 
The circled solid line shows the trajectory 
of the boson peak (for $\Delta=0$, 
its position coincides 
with the lowest van Hove singularity in the 
crystalline lattice). 
The solid line represents the upper band edge 
obtained by the CPA. 
The horizontal dotted line separates the systems 
stable in equilibrium 
from those that are mechanically unstable. 
}
\label{Phase_diagram_new}
\end{figure}

We have solved Eq.~(\ref{e3_c_8}) numerically using 
the exact energy dependence for the effective field 
for different values of disorder,  $\Delta$ 
(see Fig.~\ref{IR_new}). 
The results for the Ioffe-Regel crossover energy 
are presented in the phase diagram 
for vector vibrations in the f.c.c. 
lattice (see Fig.~\ref{Phase_diagram_new}). 
There is a minimum value 
of disorder 
($\Delta_{\rm min}\simeq 0.7$ for the disordered 
f.c.c. lattice using Eq.~(\ref{e3_c_8})), 
starting from which the 
region of the strong-scattering regime appears 
in the system. 
The strong-scattering regime becomes 
broader with increasing disorder, while 
the low-energy weak-scattering regime shrinks basically 
to zero around the critical disorder, $\Delta_*$. 
We should stress that the boundary between 
weak- and strong-scattering regimes 
should really be thought of as  a  relatively wide crossover 
strip, and Eq.~(\ref{e3_c_8}) serves for an 
order-of-magnitude estimate of the location of this strip. 
Bearing this in mind,  the high-frequency region 
at sufficiently large disorder should not properly be 
regarded as 
a region of weak-scattering because the width 
of the spectral-density 
peak there is of the same order of magnitude as 
the peak position. 
The true weak-scattering regime occurs only 
either in the low-energy regime at  
any  disorder less than the critical value, $\Delta\alt \Delta_*$, 
 or  for all energies at small disorder 
($\Delta \alt \Delta_{\rm min}$). 
The existence in these regions of small values of the 
energy and disorder  parameters   
forces the 
spectral-density width 
to be much smaller than the peak position. 
In order to illustrate how sensitive the Ioffe-Regel 
crossover energy is  to the numerical coefficients 
in Eq.~(\ref{e3_c_8}), we have 
calculated $\varepsilon^{\rm IR}_{\k\beta}$ from 
Eq.~(\ref{e3_c_8}) in which 
the spectral-density peak width, $ \Gamma_\omega$, 
has been replaced, say, by $ 2\Gamma_\omega$, thus enhancing 
the strong scattering. 
The results for this Ioffe-Regel crossover boundary 
are shown by the dot-dashed curve  
in Fig.~\ref{Phase_diagram_new}. 
It is clearly seen how the region 
 of the strong-scattering regime increases, with 
just the low-energy and 
low-disorder regions surviving for the weak-scattering 
regime. 

We have also shown in Fig.~\ref{Phase_diagram_new} 
the trajectory of the boson peak (circled solid line) 
 calculated as 
in Ref.~\cite{Taraskin_01:PRL}. 
Remarkably, it almost coincides with the trajectory 
of the Ioffe-Regel crossover for large enough disorder 
when the strong-scattering regime appears in the system. 
This feature has also been found in real glasses, e.g. 
v-SiO$_2$ \cite{Taraskin_00:PRB_IR2} and in the 
 topologically disordered IC-glass model 
(to be published elsewhere). 

The solid line in the phase diagram shows the 
upper band edge of the VDOS calculated within the CPA. 
This appears to be a very good estimate 
for the trajectory of the threshold for the 
localization-delocalization 
transition, which, in fact, we have calculated 
\cite{Taraskin_02:PhilMag} 
more exactly 
by means of a multi-fractal analysis 
\cite{Janssen_98:review} with a relatively 
high precision, $\alt 5\%$ 
to be  situated at slightly 
lower energies than the CPA band edge. 
The  CPA fails 
to reproduce the high-energy tail in the VDOS (see Fig.~\ref{VDOS}) and more 
accurate methods should be used 
\cite{Economou_83:book,Gonis_92:book}. 
It should also be mentioned that, for $\Delta > 1$, a  
low-energy  tail of localized states  
extending into the negative eigenvalue range 
appears in the true spectrum \cite{Taraskin_02:PRB_sing}. 
The CPA is not able to reproduce this effect 
for 
$1 \alt \Delta \alt \Delta_*$  
but still, for the 
majority of  states with energies far enough away from
both localization thresholds, the CPA results are very reliable.  
\subsection{Spectral density for large momentum transfer}
\label{s3_dd}

In the previous subsections, we have analysed 
the situation at small and intermediate energies of external plane waves. 
These energies correspond to a relatively small 
momentum transfer in  scattering experiments 
(see e.g.  \cite{Foret_96,Benassi_96}). 
However,  larger values of $k$ are also of interest 
both from experimental 
(see e.g. \cite{Wischnewski_98,Otomo_98,Arai_99,Nakamura_01}) 
and theoretical 
(see e.g. 
\cite{Ribeiro_98,Horbach_98:JNCS,
Taraskin_00:PRB_IR1,Taraskin_00:PRB_IR2,
Gotze_00,Ruocco_00,Simdyankin_01}) 
points of view. 
Attention has been mainly focused   
on the behaviour of the dynamic structure factor, 
$S_{\k\beta}(\omega)$, which can be 
roughly approximated in the high-temperature 
regime ($\hbar\omega_{\rm d}/T \gg 1$) by the following 
expression \cite{Ruocco_00}: 
\begin{equation}
S_{\k\beta}(\omega) \simeq 
\frac{Tk^2}{2m\hbar\omega^2} 
\sum_{\rm d} |\langle d |\k\beta\rangle|^2 
\delta(\omega-\omega_{\rm d})
\ ,  
\label{e3_dd_1}
\end{equation}
where the Debye-Waller factor has been ignored 
(cf. \cite{Taraskin_97:PRB1}). 
In Eq.~(\ref{e3_dd_1}), we have also neglected the  mixture 
of polarizations for phonons $|\k\beta\rangle$ and assume 
that $\beta$ is related to the longitudinal branch. 
In this case, the dynamical structure factor is obviously 
(see Eq.~(\ref{e2_7})) 
proportional to the spectral density, 
\begin{equation}
S_{\k\beta}(\omega) \simeq 
\frac{Tk^2}{2m\hbar\omega^2} 
A^{(\omega)}_{\k\beta} 
\equiv 
\frac{Tk^2}{m\hbar\omega} 
A_{\k\beta}(\omega^2) 
\ ,  
\label{e3_dd_2}
\end{equation}
where $ A^{(\omega)}_{\k\beta} $ is the spectral 
density in the $\omega$-representation. 

A peculiar feature has been found in the frequency dependence 
of the dynamical structure factor in 
topologically disordered glasses at large  
values of the momentum transfer. 
Namely, the appearance of a secondary peak below 
the Brillouin peak in the $\omega$-dependence 
of  $S_{\k\beta}(\omega)$ has been observed 
\cite{Horbach_98:JNCS,Gotze_00,Ruocco_00,Simdyankin_01} 
with increasing value of $k$. 
This has been related to  the boson peak 
\cite{Horbach_98:JNCS}, or interpreted as  a manifistation of  a 
microscopic relaxation process typical of  
a topologically disordered glass \cite{Ruocco_00}. 

\begin{figure}[b!] 
\centerline{\epsfig{file=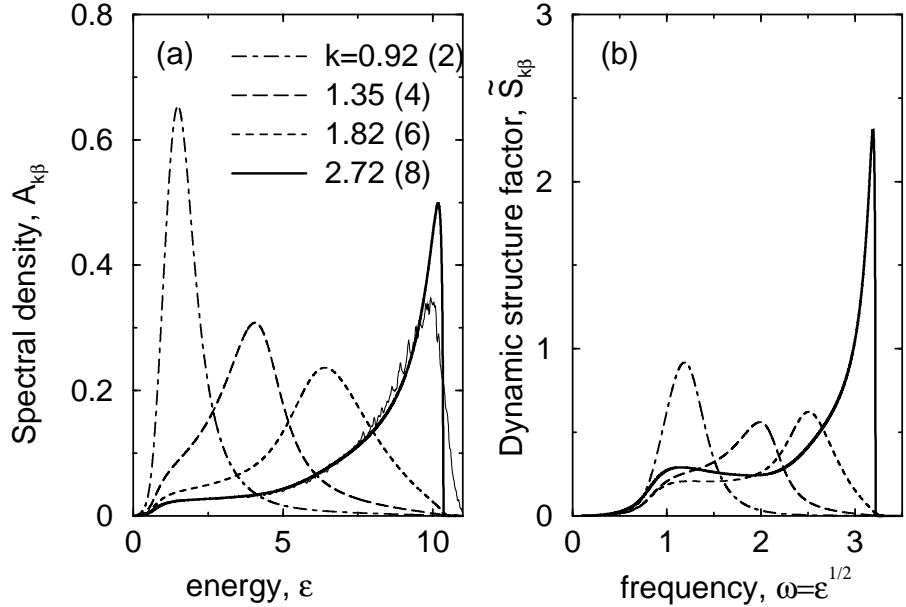,width=8truecm,angle=270}}
\vskip 20pt
\caption[]{
The evolution of the spectral density, 
$A_{\k\beta}(\varepsilon)$ (a),  and 
dynamic structure factor ${\tilde S}_{\k\beta}$ (b), 
with wavevector magnitude, $k$, as marked 
(the corresponding crystalline energies, $\varepsilon_{\k\beta}$ are given in brackets),  taken  
along the $[100]$ symmetry direction in the first 
Brillouin zone 
(the maximum value of $k=2.72$ corresponds 
to the zone boundary) for the longitudinal branch in the f.c.c.  
disordered ($\Delta = 1$) lattice.  
The thin solid line in (a) for $k=2.72$ shows the 
numerical results obtained by  KPM.  
}
\label{SD_1}
\end{figure}

 A natural question is whether or not the secondary peak 
exists for the disordered-lattice models and, if so, then 
what is the origin of this peak. 
In order to answer  these questions, first, we plot  in 
Fig.~\ref{SD_1}(a) the evolution of the spectral density 
$A_{\k\beta}(\varepsilon)$ for the longitudinal branch with 
increasing value of the wavevector, taken for 
definitness to be along the $[100]$ direction. 
It is clearly seen from this figure  that the spectral density 
has a pronounced peak (Brillouin peak) for all values of $k$ taken 
from the first Brillouin zone. 
However this peak is no longer Lorentzian for large $k$. 
It is asymmetric and has a pronounced shoulder in the 
low-energy range. 
A similar picture has been observed in  simulations 
for  topologically disordered systems. 
The Brillouin peak shifts to  higher energies, 
becomes very broad and a smooth shoulder-like 
feature appears below this peak
and may also contain  broad peak-shaped features 
\cite{Sampoli_97,Ribeiro_98,Taraskin_00:PRB_IR1,Simdyankin_01}, 
so that the spectral density starts to resemble the VDOS. 
In the low-energy regime, the spectral density goes to 
zero. 

\begin{figure}[b!] 
\centerline{\epsfig{file=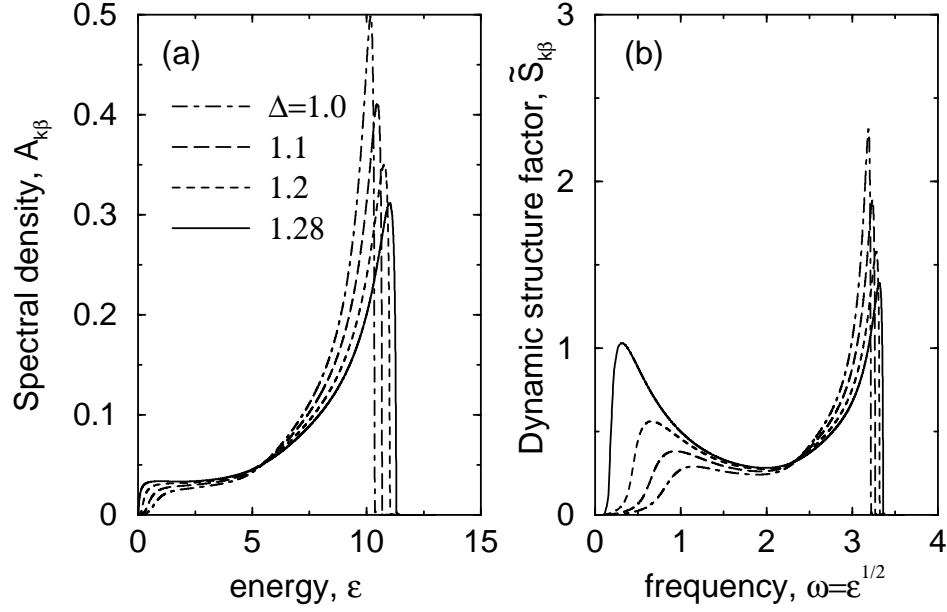,width=8truecm,angle=270}}
\vskip 20pt
\caption[]{
The evolution of the spectral density, 
$A_{\k\beta}(\varepsilon)$ (a),  and 
dynamic structure factor ${\tilde S}_{\k\beta}$ (b), 
with degree of disorder, $\Delta$, as marked 
for  $k=2.72$ (longitudinal branch)  corresponding  
to the zone boundary along $[100]$. 
}
\label{SD_Delta}
\end{figure}

The dynamical structure factor differs from the spectral 
density and contains an additional 
decaying energy (frequency) function (see Eq.~(\ref{e3_dd_2}). 
Therefore, if we plot the effective dynamic structure factor, 
$
{\tilde S}_{\k\beta}(\omega) = 
(2m\hbar/T)S_{\k\beta}(\omega) = 
(k^2/\omega^2) A^{(\omega)}_{\k\beta} = 
(k^2/2\sqrt{\varepsilon}) 
A_{\k\beta}(\varepsilon)
$, 
versus frequency (see Fig.~\ref{SD_1}(b)), 
the low-frequency features are enhanced 
due to  the $\varepsilon^{-1/2}$ contribution. 
The low-energy shoulder in the spectral density  
of the disordered-lattice models  
transforms, for large $k$ values close to the Brillouin-zone boundary,   
to the secondary peak in the dynamic structure 
factor (see the solid line in Fig.~\ref{SD_1}(b)). 
The intensity of the secondary peak depends on the degree of disorder, so that 
it becomes very pronounced around the critical disorder 
(see Fig.~\ref{SD_Delta}(b)), although it does not appear for 
the spectral density in the energy representation 
(cf Fig.~\ref{SD_Delta}(a)). 
Therefore, the presence of the secondary peak 
in $S_{\k\beta}(\omega)$ of  disordered-lattice models 
for large values of  wavevector and large degrees 
of disorder is a distinctive feature of such models. 
Consequently, the occurrence of a similar 
peak in  topologically disordered structures can 
hardly be attributed solely to topological disorder 
(cf. Ref.~\cite{Ruocco_00}).  

The origin of the secondary peak in the dynamic structure factor of 
disordered-lattice models can be understood from the 
qualitative analysis of Eqs.~(\ref{e2_18}) and 
(\ref{e3_dd_2}). 
In the low-frequency limit, $\omega\to 0$ 
(or $\varepsilon \to 0$), 
the energy dependence of the spectral 
density in Eq.~(\ref{e2_18}) is dictated 
by the imaginary part of the effective field, 
$A_{\k\beta} \propto |z''(\varepsilon)| 
\propto \varepsilon^{3/2} = \omega^3$, so that 
$S_{\k\beta} \propto A_{\k\beta}/\omega 
\propto \omega^2$. 
The low-energy limiting behaviour of 
$z''(\varepsilon)$ deviates approximately 
around the energy corresponding to the shifted 
(due to the level-repelling effect) lowest van Hove singularity, 
i.e. the first minimum in the energy 
dependence of  $z'(\varepsilon)$ 
(cf. Figs.~\ref{Z_eff}(a) and (b)). 
Above this energy, the energy dependence of the imaginary part of the 
effective field is appreciably 
weaker and basically reproduces the shape of the 
VDOS (see Fig.~\ref{Z_eff}(b)). 
This results in a weaker dependence of the 
dynamic structure factor, which shows,  
in this intermediate frequency range,  either 
the shoulder-like behaviour, if  the wavevector 
magnitude is not large enough and the 
low-frequency tail of the Brillouin peak 
plays a significant role (see the long and short dashed lines in 
Fig.~\ref{SD_1}), or a secondary peak for large 
values of $k$ (see the solid line in Fig.~\ref{SD_1}), 
when the influence of  the Brillouin peak is suppressed 
by the $\varepsilon^{-1/2}$ function in 
Eq.~(\ref{e3_dd_2}). 
Therefore, the position of the secondary peak 
in $S_{\k\beta}(\omega)$ for the disordered 
lattice models approximately coincides with 
the shifted lowest van Hove singularity and 
thus with the position of the boson peak 
\cite{Taraskin_01:PRL}. 
However, we should stress that the presence 
of the pronounced secondary peak in the 
dynamic structure factor for large values of $k$ 
does not mean a major contribution  
of plane waves with such $k$ to the boson peak, which 
is proportional to the sum of all plane waves 
(see  Ref.~\cite{Taraskin_01:PRL} for more detail). 

The $\omega^2$-dependence of the dynamic 
structure factor in the low-frequency regime found 
for the disordered-lattice models is 
consistent with a $k^4$-dependence of 
the spectral density width and is in contrast to 
the constant (non-zero), limit of 
$S_{\k\beta}(\omega)$ found 
for topologically disordered glasses 
\cite{Gotze_00,Ruocco_00,Simdyankin_01},  
and thus can be considered as a drawback of lattice models.  

\subsection{Sum-rule correlations and 
plane-wave scattering}
\label{s3_e}

We have seen in the previous subsection that, 
independent of the degree of disorder,  
 low-energy  external plane waves 
are weakly scattered in a force-constant disordered lattice. 
This is a consequence of the  
essentially different dependences 
on the plane-wave energy of the 
peak width ($ \Gamma_{\k\beta} \propto 
\varepsilon_{\k\beta}^{5/2}$)  and peak position 
($ \varepsilon_{\k\beta}^{\rm max} \propto \varepsilon_{\k\beta}$) 
of the spectral density 
(see Eqs.~(\ref{e3_2},\ref{e3_3})), 
which gives rise to the occurrence of  
weak scattering 
($\Gamma_{\k\beta} \ll \varepsilon_{\k\beta}^{\rm max}$) 
for $\varepsilon_{\k\beta} \to 0$. 
The question is how  generic is this feature. 

In order to answer this question, 
let us consider a more general formulation of the 
problem using the following Hamiltonian 
in the site representation: 
\begin{equation}
{\hat \H}= 
\sum_i(\varepsilon_i + \gamma\sum_{j\ne i}\kappa_{ij}
) 
|i\rangle\langle i| - 
\sum_{i,j\ne i}\kappa_{ij}|i\rangle\langle j| 
\ .  
\label{e3_e_1}
\end{equation}
The diagonal elements of the Hamiltonian  have two constituents: 
random on-site ''energies'', $\varepsilon_i$, 
and  the correlation contribution proportional to the sum 
of random off-diagonal elements, $\kappa_{ij}$, 
 in the same row. 
The parameter $\gamma$ controls the sum-rule 
correlations between diagonal and off-diagonal 
elements. 
If $\gamma=1$ and $\varepsilon_i=0$,  the Hamiltonian 
(\ref{e3_e_1}) describes the scalar vibrational problem, 
while the case $\gamma=0$ corresponds to the 
well-known electron  Anderson Hamiltonian with on-site and 
off-diagonal disorder \cite{Economou_83:book}.  
The spectral properties of a similar Hamiltonian,  
but defined on a positionally  random set of sites with 
deterministic transfer integrals,   have been studied in 
Ref.~\cite{Mezard_99}.

The scattering properties of external plane waves 
in the disordered structure described by the 
Hamiltonian (\ref{e3_e_1}) can be studied  
using  spectral densities for which 
the mean-field approach works very well 
(as shown above for the vibrational problem). 
It is easy to show that only in the case of 
$\gamma=1$ and $\varepsilon_i=0$ can  the 
Hamiltonian (\ref{e3_e_1})  be reduced 
to the form of Eq.~(\ref{e1_8}) which has  
a multiplicative effective mean field 
(the energy of the quasiparticles is the product 
of the effective field and the bare energy, 
${\tilde\varepsilon}_{\k\beta}  =  z(\varepsilon)\varepsilon_{\k\beta}$ -  
 see Eq.~(\ref{e2_13_3})). 
For $\gamma\ne 1$, the situation is quite different, 
because now the effective field has an additive constituent as well, 
and this drastically changes the scattering properties of 
plane waves. 
Below, we demonstrate this, considering the simplest 
particular case for the Hamiltonian (\ref{e3_e_1}) 
characterized by $\gamma=0$, $\kappa_{ij}=\kappa_0$ 
and random $\varepsilon_i$ taken from the box distribution 
of half-width, $\Delta^{\rm el}$, 
centred around $\varepsilon_0=0$, i.e. 
the standard Anderson Hamiltonian with diagonal disorder 
\cite{Kramer_93}, for the simple cubic lattice. 

\begin{figure}[b!] 
\centerline{\epsfig{file=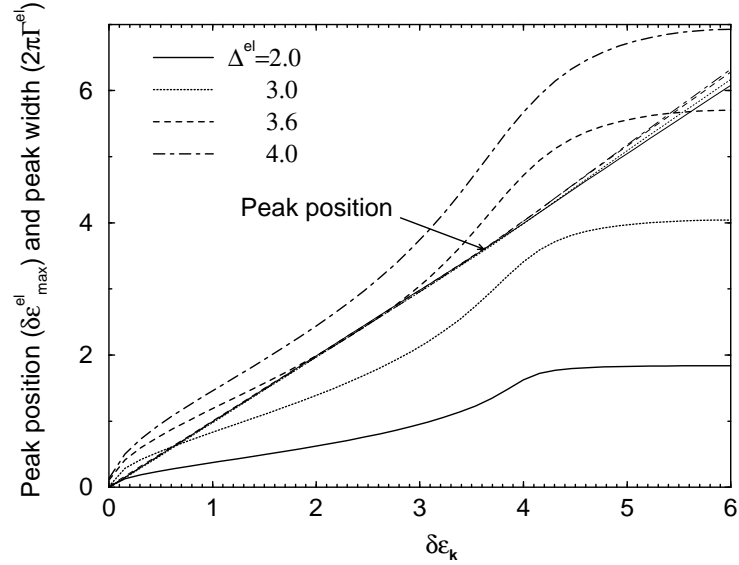,width=8truecm,angle=270}}
\vskip 20pt
\caption[]{
The peak positions, $  \delta\varepsilon^{\rm el}_{\rm max} $ 
(thin lines), and peak widths, 
$2\pi\Gamma^{\rm el}$ (thick lines) for the 
electron spectral densities 
for the Anderson Hamiltonian with on-site disorder 
for the simple cubic lattice, versus 
plane-wave energy, $\delta\varepsilon_{\k}$, 
at different values of disorder as marked. 
The crossing points of the curves for peak positions and peak widths 
correspond to the Ioffe-Regel crossover energy for electrons. 
}
\label{IR_electrons}
\end{figure}

The CPA spectral density, $A^{\rm el}_{\k}(\varepsilon) = 
-(\pi)^{-1}{\rm Im}(\varepsilon - {\tilde\varepsilon}_\k)^{-1}$, 
for the Anderson Hamiltonian can be expressed 
via the quasiparticle energy, 
${\tilde\varepsilon}_\k = \varepsilon_\k + 
\tilde\varepsilon(\varepsilon)$, which is now 
the sum of the crystalline energy, $\varepsilon_{\k}$, 
and the effective field, 
$\tilde\varepsilon(\varepsilon)=\tilde\varepsilon'(\varepsilon) 
+ i \tilde\varepsilon''(\varepsilon)$. 
The spectral densities  $ A^{\rm el}_{\k}(\varepsilon)$ 
are peak shaped 
and in the low-energy limit, 
$ \delta\varepsilon_\k = (\varepsilon_\k - 
 \varepsilon_{\rm min}^{\rm cryst})  \to 0$ 
 (the energy 
$\delta\varepsilon_\k $ is referred to the lower 
crystalline band edge, 
$\varepsilon_{\rm min}^{\rm cryst}=-\kappa_0 Z$,  
where $Z$ is the coordination number), 
the position and width of the peak 
can be estimated from the following relations: 
\begin{equation}
\delta\varepsilon^{\rm el}_{\rm max} \sim 
\delta\varepsilon_\k
\ , 
\label{e3_e_6} 
\end{equation}
and 
\begin{equation}
\Gamma^{\rm el}(\varepsilon_\k) \simeq 
\frac{1}{2}\pi\chi^{\rm el} (\Delta^{\rm el})^2 
\sqrt{\delta\varepsilon^{\rm el}_{\rm max}} 
\propto 
(\Delta^{\rm el})^2 
\sqrt{\delta\varepsilon_\k}
\ ,  
\label{e3_e_7} 
\end{equation}
where the peak position, 
$ \delta\varepsilon^{\rm el}_{\rm max} = 
(\varepsilon^{\rm el}_{\rm max} - 
\varepsilon_{\rm min}^{\text{band}})$, 
is measured from the 
CPA lower band boundary, $\varepsilon_{\rm min}^{\rm band}$ 
(see Fig.~\ref{IR_electrons}). 
The last expression is a consequence  of 
the fact  that 
$ 
\mbox{Im}\left[G^{\rm cryst}
               (\varepsilon-\tilde\varepsilon) 
         \right]
\simeq 
-\pi 
 \chi^{\rm el}\sqrt{\varepsilon - 
\varepsilon_{\rm min}^{\rm cryst} }$, 
if 
$ \left(\varepsilon - 
\varepsilon_{\rm min}^{\rm cryst}\right)  \to 0$ with 
$\chi^{\rm el}$ being the constant coefficient similar 
to the coefficient $\chi_{\rm Deb}$ in Eq.~(\ref{e22_1a}). 

\begin{figure}[b!] 
\centerline{\epsfig{file=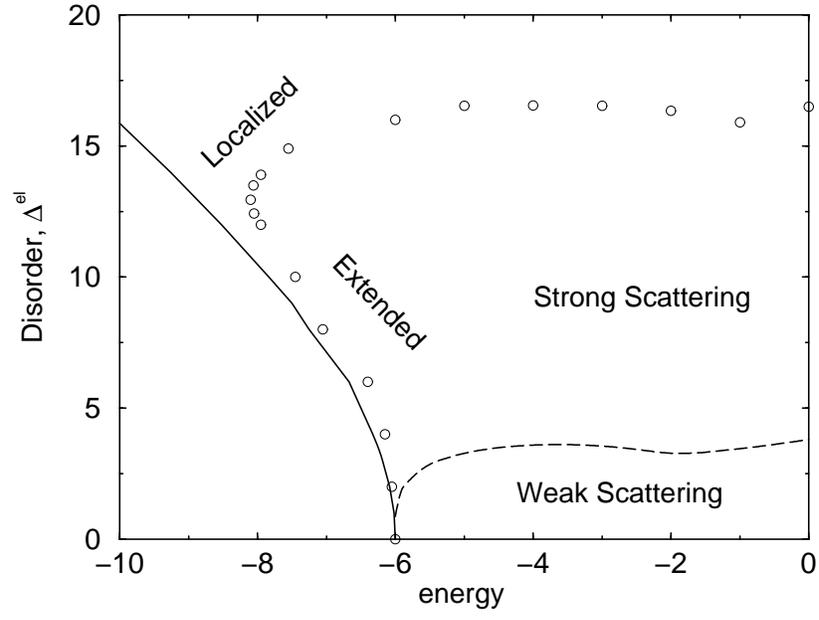,width=8truecm,angle=270}}
\vskip 20pt
\caption[]{
Phase diagram for electrons described by 
the Anderson model with on-site energy disorder 
 in the simple cubic  
lattice. 
The dashed line corresponds to the 
Ioffe-Regel crossover plane-wave energy 
obtained from Eq.~(\protect\ref{e3_e_8}). 
The solid line represents the lower  band edge 
obtained by the CPA and the  circles show the 
trajectory for the localization-delocalization 
threshold \protect\cite{Bulka_87}.  
}
\label{Phase_diagram_electrons}
\end{figure}

In full analogy to vibrations (see Eq.~(\ref{e3_c_8})), 
the Ioffe-Regel 
crossover between weak- and strong-scattering 
regimes for electrons can be defined  in terms of 
 the following condition: 
\begin{equation} 
\frac{ 
\delta\varepsilon^{\rm el}_{\rm max} 
(\varepsilon_\k)
}
{2\pi} 
\sim 
\Gamma^{\rm el}(\varepsilon_\k) 
\ . 
\label{e3_e_8}
\end{equation}
In contrast to vibrations, the spectral-density peak width 
for electrons scales more slowly  
 than the peak 
position as a function of the plane-wave energy  
(see Eqs.~(\ref{e3_e_6})-(\ref{e3_e_7}) and  
Fig.~\ref{IR_electrons}). 
This means that, in the long-wavelength limit, 
$\delta\varepsilon_\k \to 0$, the width of 
the spectral density 
is always larger than the peak position, and 
the strong-scattering regime occurs for plane waves having 
energies around 
the lower crystalline  band edge 
(see  Fig.~\ref{Phase_diagram_electrons}). 
The situation is opposite for vibrations, 
for which $\Gamma_{\k\beta} 
\propto \varepsilon_{\k\beta}^{5/2}$ 
(cf. Eqs.~(\ref{e3_3}) and (\ref{e3_e_7})). 
As a consequence,   
the  shapes of the regions associated with 
the weak- and strong-scattering 
regimes for plane waves in the phase diagram 
are different for electrons and vibrations 
(cf. Figs.~\ref{Phase_diagram_new} and 
\ref{Phase_diagram_electrons}). 
Now, for electron plane waves with energies around 
the lower crystalline band edge, the scattering is always strong 
and the weak-scattering regime is realized only in the midband 
range for sufficiently small disorder. 
In contrast to vibrations, only a single  parameter, 
the strength of disorder, $\Delta^{\rm el}$, rather than 
two parameters, the degree of disorder $\Delta$ and the energy, 
$\varepsilon_{\k\beta}$, for vibrations, 
controls the appearance of the  weak-scattering regime.  
The boundary between the weak- and strong-scattering 
regimes (see the dashed line in Fig.~\ref{Phase_diagram_electrons}) 
is again not precise and should 
 be thought  of as  an estimate for the crossover energy. 

It should be mentioned that the electron phase diagram 
presented in Fig.~\ref{Phase_diagram_electrons}, 
for the case of very small disorder ($\Delta^{\rm el} \to 0$),  
supports 
 a standard concept for electrons \cite{Mott_79:book}, 
according to which the Ioffe-Regel crossover 
from the weak- to the strong-scattering regime 
corresponds to the transition from  extended 
states to  localized ones. 
Indeed, the Ioffe-Regel crossover boundary (the dashed line 
in Fig.~\ref{Phase_diagram_electrons}) almost 
coincides with the localization-delocalization 
 threshold (the circles in 
Fig.~\ref{Phase_diagram_electrons}) \cite{Bulka_87} 
and the CPA band edge (solid line) as $\Delta^{\rm el} \to 0$. 
At high disorder, the localization-delocalization 
transition, however,  is not associated with the 
Ioffe-Regel crossover, 
because all plane waves for such  disorder are 
strongly scattered. 

In the more general case of diagonal and off-diagonal 
disorder  coexisting in  the Anderson Hamiltonian (but 
with $\gamma=0$), the most successful 
homomorphic cluster CPA 
\cite{Yonezawa_79,Li_88} results in an effective 
field which  necessarily has  an additive constituent 
and the above conclusions still hold. 
In the case of $0<\gamma<1$, the mean field can be   
constructed as well but it also has an additive 
constituent which disappears only for $\gamma=1$ 
(to be published elsewhere). 
Thus,  we can conclude that the existence of the low-energy 
weak-scattering regime is characteristic only of the vibrational 
problem  and it results from the exact sum-rule 
correlations in the corresponding Hamiltonian. 

%
%
%
%
%
%
\section{Conclusions}
\label{s4}

We have analytically  investigated 
 classical harmonic atomic vibrations in disordered lattices. 
The f.c.c. atomic lattice  with force-constant disorder 
described by a  box probability distribution 
has been chosen for analysis. 
The distributions of matrix elements in diagonal 
and off-diagonal blocks in the dynamical matrix are rather similar 
to those of topologically disordered models. 
The atomic  vibrations were treated as vectors and 
analysed within the framework of the single-bond 
coherent potential approximation (CPA); 
the results are essentially the same as those found from a precise   
numerical study. 
The CPA self-consistency equation has been derived 
for vector models and solved analytically in the limiting 
cases of small disorder and low energies. 

The CPA solution of the problem has allowed us to investigate 
the spectral properties of the model in terms of 
spectral densities. 
We have shown 
that the Rayleigh law for the spectral-density width 
is intrinsic for  vector vibrations in disordered 
lattices. 

We have also investigated the Ioffe-Regel crossover 
between weak and strong scattering for vector vibrations. 
The  regime of weak scattering 
takes place at all energies for small degrees of disorder, but  
in the low-energy regime occurs only for large 
disorder. 
The existence of a disorder-independent weak-scattering regime 
for low-energy external plane waves  
is a consequence of the sum-rule correlations in the 
dynamical matrix, typical for the vibrational problem 
only. 
We have demonstrated 
that the Ioffe-Regel 
 crossover occurs in the boson-peak region and is not  
related to the localization-delocalization 
transition 
for vector vibrations in the disordered lattice studied. 
In contrast, the Ioffe-Regel crossover for 
electrons (described by the Anderson model  
with on-site energy disorder in the simple cubic lattice) 
exists only at low values of disorder, where it {\it is} 
associated with the localization-delocalization transition. 
At higher values of disorder for the Anderson model, 
the transition is 
between strongly-scattered extended states and 
localized states. 

We have found that a secondary peak appears in the dynamic structure factor 
below the Brillouin peak for sufficiently large values of $k$ 
for the lattice models as previously found in simulations 
of topologically disordered glasses. 
Hence this feature cannot be due solely to the presence 
of topological disorder, 
nor can it be due to relaxations. 
%
%
%
%

%
%
%
\section*{Appendix A}
\label{app_A}
Here we give  the derivation of the self-consistency 
CPA equation for vector vibrations. 
It is known \cite{Webman_81,Odagaki_81} that the self-consistency equation 
for the effective force constant, 
$\tilde\kappa(\varepsilon)$, 
in the single-bond approximation can be written 
as follows: 
\begin{equation}
\left \langle
\frac{\delta\kappa}
{
1-\delta\kappa 
\langle ij| {\tilde {\hat\G}}| ij \rangle 
}
\right\rangle
|ij\rangle\langle ij|=0 
\ ,    
\label{e2_13}
\end{equation}
where $ \delta\kappa=\kappa_{ij}-\tilde\kappa $ and 
the effective Green's function, 
$ {\tilde {\hat\G}} = \left[ \varepsilon - 
{\tilde {\hat\H}} \right]^{-1} $, is defined for the effective Hamiltonian, 
$ 
{\tilde {\hat\H}} =
 (\tilde\kappa/2) \sum\limits_{i,j} 
|ij\rangle\langle ij| 
$.  
Our aim is to find the matrix element, 
$\langle ij|{\tilde {\hat\G}}| ij \rangle$, for vector vibrations. 
This can be rewritten as 
\begin{equation}
\langle ij|{\tilde {\hat\G}}| ij \rangle = 
2\sum_{\alpha,\alpha'} ({\hat\r}^{(0)}_{ij})_{\alpha}
({\hat\r}^{(0)}_{ij})_{\alpha'}
\left( {\tilde G}_{i\alpha,i\alpha'}-
{\tilde G}_{i\alpha,j\alpha'}
\right)
\ ,    
\label{e2_13_1}
\end{equation}
and this can then  be expressed in terms of diagonal elements 
of the Green's function in the site basis 
for the reference crystal. 
In order to show this, let us first 
rewrite the matrix elements, 
$ 
{\tilde G}_{i\alpha,j\alpha'}
$ 
in the plane-wave basis, $\{|\k,\beta\rangle\}$, 
\begin{equation} 
{\tilde G}_{i\alpha,j\alpha'} = 
\frac{1}{N} \sum_{\k\beta}
(\n_\beta)_\alpha  (\n_\beta)_{\alpha'}
\frac{ 
\exp\left\{i\k(\r_i^{(0)}-\r_j^{(0)})\right\} 
}
{
\varepsilon - {\tilde\varepsilon}_{\k\beta}
}
\ ,      
\label{e2_13_5}
\end{equation}
where $\n_\beta$ is a $D$-dimensional unit 
polarization 
vector for the branch $\beta$ 
 and the effective energy, 
$ {\tilde\varepsilon}_{\k\beta}$,   
obeys the relation: 
\begin{equation}
{\tilde\varepsilon}_{\k\beta}
=  z(\varepsilon)\varepsilon_{\k\beta} = 
\tilde\kappa \sum_{j(i)}
\left(
\r_{ij}^{(0)}
\n_\beta
\right)^2 
\left( 
1-\exp\{i\k\r_{ij}^{(0)}\}
\right) 
\ ,    
\label{e2_13_3}
\end{equation}
with 
$z(\varepsilon)={\tilde \kappa}(\varepsilon)/\kappa_0$ 
being the dimensionless effective 
force constant and  
the index $j(i)$ running over all nearest 
neighbours to the site $i$, i.e. $j=1,\dots Z$, with 
$Z$ standing for the number of nearest neighbours  
 (all the sites in a crystal are 
equivalent and $\varepsilon_{\k\beta}$, in fact, 
does not depend on $i$). 

Substituting  expression (\ref{e2_13_5})  back 
into Eq.~(\ref{e2_13_1}) we obtain: 
\begin{equation} 
\langle ij|{\tilde {\hat\G}}| ij \rangle = 
\frac{2D}{Z{\tilde\kappa}}
\left(
\varepsilon {\tilde G}(\varepsilon)
-1 \right)
\ ,    
\label{e2_13_6}
\end{equation}
where 
\begin{equation} 
{\tilde G}(\varepsilon) \equiv 
{\tilde G}_{ii}(\varepsilon) \equiv 
\frac{1}{D}\sum_\alpha 
{\tilde G}_{i\alpha,i\alpha}(\varepsilon)
=
\frac{1}{ND}\sum_{\k\beta}
\frac{1}{\varepsilon-{\tilde\varepsilon}_{\k\beta}}
\      
\label{e2_13_7}
\end{equation}
is the average on-diagonal (independent of $i$) 
element of the effective Green's 
function in the site basis. 
The effective Green's function, ${\tilde G}(\varepsilon)$, 
can be expressed via the crystalline one, 
$ G^{\rm cryst}(\varepsilon)$, as 
\begin{equation} 
 {\tilde G}(\varepsilon)  = 
\frac{1}{z} G^{\rm cryst}(\varepsilon/z)
\   .    
\label{e2_13_8}
\end{equation}
Finally, the expression for 
$ 
\langle ij|{\tilde {\hat\G}}| ij \rangle 
$ 
reads: 
\begin{equation}
\langle ij|{\tilde {\hat\G}}| ij \rangle  =
\frac{2D}{Zz\kappa^{(0)}}
\left[\frac{\varepsilon}{z}G^{\rm cryst}
\left(\frac{\varepsilon}{z}\right) -1 
\right]
\ ,      
\label{e2_14}
\end{equation}
and this solves the problem when substituted back 
into Eq.~ (\ref{e2_13}). 

The dimensionless effective 
spring constant 
$z(\varepsilon)=\tilde\kappa/\kappa_0 $  
 fully determines  the effective mean field.  
This is a complex quantity  
and should be found from the solution 
of the self-consistent equation (\ref{e2_13}). 
The self-consistent equation (\ref{e2_13}) for vector vibrations 
is very similar to that for the scalar models 
\cite{Webman_81,Odagaki_81,Schirmacher_98,Martin_Mayor_00}. 
The only difference is the  factor $D$ in the denominator of 
Eq.~(\ref{e2_13}),  with $\langle ij|{\tilde {\hat\G}}| ij \rangle$ 
given by Eq.~(\ref{e2_14}). 
For scalar models, this is absent, even for $D$-dimensional lattices 
with $D\ne 1$. 

The crystalline Green's function, 
$G^{\rm cryst}(x)$,  
with  complex argument $x=\varepsilon/z$, can 
be easily found as an analytical 
continuation from the real axis of its imaginary part 
(VDOS) \cite{Ehrenreich_76}.  
After that, the self-consistent equation can be solved 
numerically in the general case, and analytically 
in some limiting cases (e.g. $\varepsilon \to 0$ and/or 
$\Delta \to 0$), and the effective 
field can be found.  

If  $z(\varepsilon)$ is known,  
the task of evaluating the spectral density becomes 
straightforward. 
Indeed, 
bearing in mind that the energy, 
$ {\tilde\varepsilon}_{\k\beta} $, of  the eigenstates  
in the effective crystal is connected with the energy, 
$\varepsilon_{\k\beta}$,  of 
the similar eigenstates in the reference crystal  according 
to  relation (\ref{e2_13_3}),  
it is easy to see that the VDOS and diagonal 
elements of the spectral-density 
operator in the crystalline plane-wave basis 
satisfy Eqs.~(\ref{e2_17})-(\ref{e2_18}).  

%
%
%
%
%
%
%
%
%
%

%
%
%
\section*{Appendix B}
\label{app_B}

Here we show how 
the self-consistent Eq.~(\ref{e22_1}) 
can be solved analytically in the 
low-energy limit, $\varepsilon \to 0$, and how 
the effective field, $z(\varepsilon)$, can be found in this case. 
In order to do this, first  
we find the expression for the function  
$\alpha(\varepsilon)=\alpha'(\varepsilon) + 
i\alpha''(\varepsilon)$ given by Eq.~(\ref{e22_1aa}) 
via $z(\varepsilon)$ and $\varepsilon$. 
This can be done, bearing in mind the following 
relations for the crystalline Green's function, 
$ G^{\rm cryst}(\varepsilon/z) $, 
in the low-energy limit, when the Debye approximation 
for the crystalline VDOS, 
\begin{equation}
 g^{\rm cryst}\simeq 
\chi_{\rm Deb}\varepsilon^{(D-2)/2} \ , 
\label{e22_1a} 
\end{equation}
 is valid: 
\begin{equation}
\mbox{Re} 
\left[
      G^{\rm cryst}
      \left(
            \frac{\varepsilon}{z}
      \right)
\right]
\simeq 
\mbox{const}\ \ \ \ \mbox{and} \ \ \ \ \mbox{Im}
\left[
      G^{\rm cryst}
      \left(\frac{\varepsilon}{z}
      \right)
\right] 
\simeq 
-\pi\chi_{\rm Deb}
\left(\frac{\varepsilon}{z'}
\right)^{\frac{D-2}{2}}
\ . 
\label{e22_2} 
\end{equation}
Substituting Eq.~(\ref{e22_2}) into 
Eqs.~(\ref{e22_1aa}) and (\ref{e2_14}), 
we obtain the leading terms in $\varepsilon$ for 
$\alpha$: 
\begin{equation}
\alpha'(\varepsilon) \simeq -\frac{2D}{Zz'}
\ , 
\label{e22_3} 
\end{equation}
and 
\begin{equation}
\alpha''(\varepsilon) \simeq -\frac{2D}{Z(z')^2}
\left[ 
      \frac{\pi\chi_{\rm Deb}}{z'}
          \left(
                \frac{\varepsilon}{z'}
          \right)^{\frac{D}{2}}-z''
\right] 
\ . 
\label{e22_4} 
\end{equation}
These expressions can then be used for an expansion 
of the logarithm in Eq.~(\ref{e22_1}) in terms of 
$\varepsilon$, which, after some algebra, 
 results in the following 
final relations for $z'(\varepsilon)$ and 
$z''(\varepsilon)$: 
\begin{equation}
f(z') \equiv 
\frac{2\Delta}{z'(\gamma+1)} - \mbox{ln} 
\left[
\frac{1+\Delta+\gamma z'} 
{1-\Delta + \gamma z'} 
\right] = 0
\ ,  
\label{e22_5} 
\end{equation}
and 
\begin{equation}
z''(\varepsilon) \simeq 
\frac{2\pi D \chi_{\rm Deb}}{Zz'}
\left( 
     \frac{\varepsilon}{z'} 
\right)^{\frac{D}{2}}
\left\{ 
      \left[ 
            \Delta^2 +\left( (\gamma +1)z'\right)^2 - 
             \left(1+\gamma z'\right)^2 
     \right]^{-1} - 
     \left[
          (\gamma+1)(z')^2
     \right]^{-1} 
\right\}^{-1} 
\ , 
\label{e22_6} 
\end{equation}
where $\gamma \equiv (Z/2D)-1$. 

In fact, relation (\ref{e22_5}) is the equation to find 
the real part of the effective force constant, $z'$. 
The important point is that all the parameters entering 
this equation are energy 
independent, so that $z'(\varepsilon \to 0) 
\simeq z'(0) = \mbox{const}$. 
Eq.~(\ref{e22_5})  can be solved numerically 
in the  case of arbitrary disorder and 
analytically  in the 
small-disorder limit. 

The imaginary part of the effective force constant, 
as follows from Eq.~(\ref{e22_6}),  
scales with energy as $z''(\varepsilon) \propto 
\varepsilon^{D/2}$. 
Such a behaviour is basically a consequence 
of the Debye law for the VDOS in the low-energy 
regime (see Eq.~(\ref{e22_1a})) 
and, in this sense, is general for disordered lattices. 
This dependence also determines the scaling behaviour 
of the spectral-density peak width with energy, the 
well-known Rayleigh law in this case. 

Eqs.~(\ref{e22_5})-(\ref{e22_6}) 
have   important consequences for  
the energy dependence of the VDOS in the low-energy limit. 
Indeed, in the limit $\varepsilon \to 0$, 
the expression (\ref{e2_17}) 
for the VDOS can be rewritten in the form 
of Eq.~(\ref{e22_7}). 

In the low-disorder limit, $\Delta \to 0$, Eq.~(\ref{e22_5}) 
can be solved with respect to $z'$ 
while  Eq.~(\ref{e22_6})  can be simplified,  so that
\begin{equation}
z'(\varepsilon) \simeq 1-\frac{2D}{3Z}\Delta^2 
(1+O(\varepsilon))
\  ,    
\label{e22_9} 
\end{equation}
\begin{equation}
z''(\varepsilon) \simeq -\frac{2\pi\chi_{\rm Deb}D}{3Z}
\varepsilon^{D/2}\Delta^2
\  .   
\label{e22_10} 
\end{equation}
These relations are used  in deriving  Eq.~(\ref{e22_11}).  


\newpage

------------------------------------------------------------------------

\end{document}